\documentclass{jfm}
\usepackage{graphicx,natbib}

\ifCUPmtlplainloaded \else
  \checkfont{eurm10}
  \iffontfound
    \IfFileExists{upmath.sty}
      {\typeout{^^JFound AMS Euler Roman fonts on the system,
                   using the 'upmath' package.^^J}%
       \usepackage{upmath}}
      {\typeout{^^JFound AMS Euler Roman fonts on the system, but you
                   dont seem to have the}%
       \typeout{'upmath' package installed. JFM.cls can take advantage
                 of these fonts,^^Jif you use 'upmath' package.^^J}%
      }
  \else
  \fi
\fi

\ifCUPmtlplainloaded \else
  \checkfont{msam10}
  \iffontfound
    \IfFileExists{amssymb.sty}
      {\typeout{^^JFound AMS Symbol fonts on the system, using the
                'amssymb' package.^^J}%
       \usepackage{amssymb}%
       \let\le=\leqslant  
       \let\ge=\geqslant  
      }{}
  \fi
\fi

\ifCUPmtlplainloaded \else
  \IfFileExists{amsbsy.sty}
    {\typeout{^^JFound the 'amsbsy' package on the system, using it.^^J}%
     \usepackage{amsbsy}}
    {}
\fi

\newsavebox{\astrutbox}
\sbox{\astrutbox}{\rule[-5pt]{0pt}{20pt}}

\title[Emergent order in rheoscopic swirls]{Emergent order in rheoscopic swirls}

\author[M. Wilkinson, V. Bezuglyy and B. Mehlig]
{Michael Wilkinson${^1}$, Vlad Bezuglyy$^{1,3}$ and Bernhard Mehlig${^2}$}

\affiliation{ $^{1}$Department of Mathematics and Statistics, The Open
University, Walton Hall,
Milton Keynes, MK7 6AA, England. \\[\affilskip]
$^{2}$Department of Physics, G\"oteborg University, 41296
Gothenburg, Sweden. \\[\affilskip]
${^3}$Current address: Nancy-Universit\'es,
LAEGO, rue du doyen Roubault, 54501, Vandoeuvre-les-Nancy, France.
\\}

\pubyear{}
\volume{}
\pagerange{}
\date{??}
\begin{document}

\maketitle

\begin{abstract}
We discuss the reflection of light by a rheoscopic fluid (a suspension of microscopic rod-like crystals) in a steady two-dimensional flow. This is determined by an order parameter which is a non-oriented vector, obtained by averaging solutions of a nonlinear equation containing the strain rate of the fluid flow. Exact solutions of this equation are obtained from solutions of a linear equation which are analogous to Bloch bands for a one-dimensional Schr\"odinger equation with a periodic potential. On some contours of the stream function, the order parameter approaches a limit, and on others it depends increasingly sensitively upon position. However, in the long-time limit a local average of the order parameter is a smooth function of position in both cases. We analyse the topology of the order parameter and the structure of the generic zeros of the order parameter field.
\end{abstract}

\section{Introduction}
\label{sec: 1}

\begin{figure}
\centerline{\includegraphics[width=11cm]{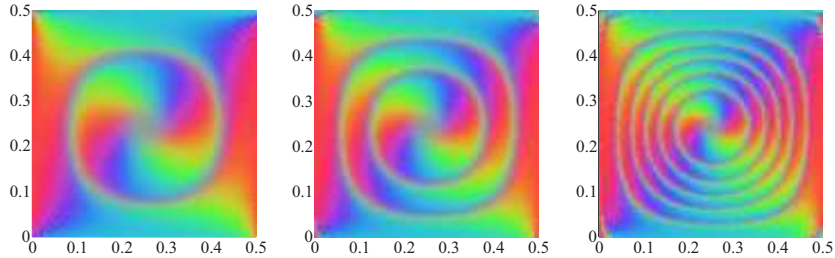}}
\caption{\label{fig: 1} Illustrates a spiral pattern which may be generated by motion of a rheoscopic fluid in a two-dimensional cellular flow, with stream function $\psi(x,y)=\sin(x)\sin(y)/2\pi$. The flow is visualised by reflected light from three different coloured sources, as shown schematically in figure \ref{fig: 2} and as described in detail in section \ref{sec: 3}. The arms of the spiral tighten as time increases. The times are (successively from left to right) $t=5$, $t=9$, and $t=18$, and the parameters in (\ref{eq: 2.1}), (\ref{eq: 2.2}) are $\alpha_1=0.95$, $\alpha_2=0.05$.}
\end{figure}

\begin{figure}
\centerline{\includegraphics[width=9cm]{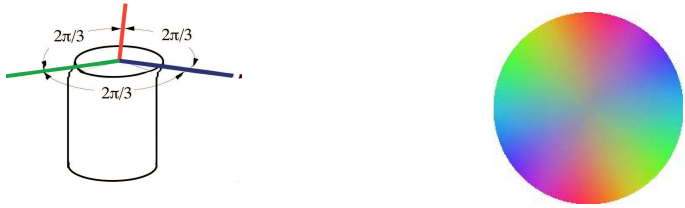}}
\caption{\label{fig: 2} {\bf a} The direction and degree of ordering of the axes of the crystals in a rheoscopic fluid can be revealed by scattering light from red, green and blue sources arranged around the sample. {\bf b} The degree of order of the particles is described by an order parameter vector $\mbox{\boldmath$\zeta$}$ lying within a unit circle, which points in the predominant direction of alignment, with the magnitude $0\le\vert\mbox{\boldmath$\zeta$}\vert\le 1$ indicating the degree of alignment. {\bf c} The colour of the scattered light is a function of the order parameter: because the orientation of the vector $\mbox{\boldmath$\zeta$}$ is irrelevant, this colour map is symmetric under reflection.
}
\end{figure}

Rheoscopic fluids are suspensions of microscopic rod-like crystals which are brought into alignment by a fluid flow. They enable the flow to be visualised due to the angular dependence of light reflection from the crystals \cite[]{Mat+84} and similar suspensions are used to make art installations \cite[]{Ree02} and to enhance the appearance of cosmetic products \cite[]{Pre84}. The patterns produced by rheoscopic agents bear a complex relation to the underlying flow, which is not yet thoroughly understood. For example, a simple stirring motion produces an increasing tightly wound spiral pattern, illustrated in figure \ref{fig: 1} (which shows additive mixing of light scattered from red, green and blue sources, as illustrated in figure \ref{fig: 2}). {\sl A priori}, it is not clear how the orientation of the rods should depend upon position within this pattern, or how it will evolve in the long-time limit. In this paper we motivate the definition of an order parameter for the alignment of the crystals in two-dimensional flows, and show how it may be related to the colour of the reflected light. We use this to analyse the long-time limit of the orientation patterns formed by steady flows in two dimensions, where trajectories of the small crystals follow contours of the stream function, $\psi(x,y)$.

\begin{figure}
\centerline{\includegraphics[width=6cm]{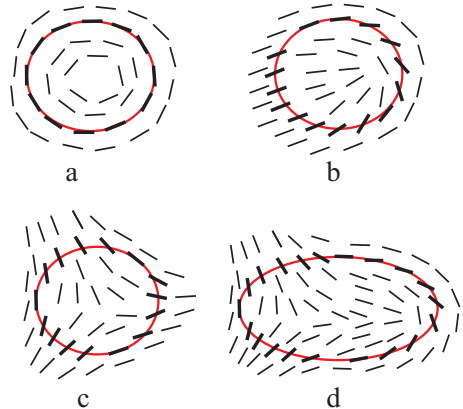}}
\caption{\label{fig: 3} The Poincar\' e index is a topological invariant. For a vector field in the plane, the Poincar\' e index of a closed curve is the number of $2\pi$ clockwise rotations of the vector field as the curve is traversed, also clockwise. Curves with a non-zero Poincar\'e index must encircle a singularity of the field. Because the axial vector of the rod-like crystals is non-oriented, singularities with half-integer Poincar\'e index are possible: {\bf a} is a vortex, {\bf b} is a {\em core} and {\bf c} is a {\em delta}, with indices $+1$, $\frac{1}{2}$, $-\frac{1}{2}$ respectively. If the curve encloses two singularities, their indices are added: for example the Poincar\'e index of the curve in {\bf d} is $\frac{1}{2}-\frac{1}{2}=0$.
}
\end{figure}

The motion of small ellipsoidal bodies in steady two-dimensional flows was previously considered by \cite{Sze93}, who showed that there may exist orbits (closed contours of $\psi(x,y)$) where the principal axis approaches a constant direction, as well as orbits where the axis tumbles and where spiral patterns such as that in figure \ref{fig: 1} are seen. He also showed that the regions where alignment occurs are characterised by a topological index, which was termed the \lq flip number', but which is in fact equal to twice the Poincar\'e index (the definition of the Poincar\'e index is illustrated in figure \ref{fig: 3}). His results raise a variety of interesting questions concerning the textures of rheoscopic flows, which are resolved in this paper. What is the actual appearance of the system under reflected light? In the regions where the crystals tumble, they may still have a preferred alignment. Does their alignment approach a time-independent limit in the regions where the crystals tumble, and if so, how is this limit approached? Is there an abrupt change in appearance on crossing from the tumbling region to the aligning region? These questions are most directly addressed by analysing an order parameter vector $\mbox{\boldmath$\zeta$}(\mbox{\boldmath$r$},t)$ which describes the predominant direction of alignment of the crystals, even in regions where they are tumbling.

We show that in the tumbling regions the order parameter forms a progressively more tightly-wound spiral pattern, having an increasingly sensitive dependence upon position. In the long-time limit the order parameter will fluctuate on a scale which is below the resolving power of the eye, and it is necessary to consider a local average of the orientations within  a small disc, from which we determine an averaged order parameter, denoted by $\langle \mbox{\boldmath$\zeta$}\rangle$. We show how the smoothly-varying local average is calculated, so that in the long-time limit the ordering of the rods is described by a smoothly varying function $\langle\mbox{\boldmath$\zeta$}\rangle(\mbox{\boldmath$r$})$, defined in both the tumbling and the aligning regions. We find that this function has no discontinuity at the boundary between the regions.
This raises a further question. It is observed that the aligning regions have different Poincar\'e indices, which implies that $\langle \mbox{\boldmath$\zeta$}\rangle(\mbox{\boldmath$r$})$ must have some form of singularity in the tumbling region. What is the form of these singularities? We identify the normal forms for generic singularities which are nodal points of the field $\langle \mbox{\boldmath$\zeta$}\rangle$. These singularities have Poincar\'e index $\pm\frac{1}{2}$ and have structures which are analogous to singularities which are seen in the ridge patterns of fingerprints.

\begin{figure}
\centerline{\includegraphics[width=10cm]{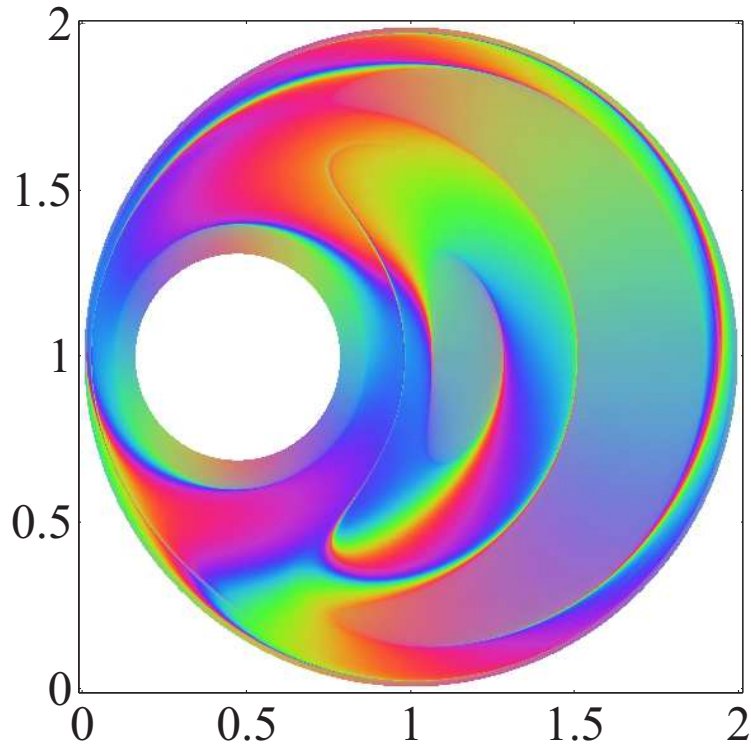}}
\caption{\label{fig: 4} Illustrating the reflection of red, green and blue light from rheoscopic fluid in a \lq journal bearing' flow in the long-time limit. This is a steady two-dimensional flow between two non-concentric rotating non-slip circular walls. In this example both walls rotate in the same direction, with the angular velocity of the inner boundary exceeding that of the outer boundary by a factor of $20$. The aspect ratio of the elliptical particles is $\beta=\sqrt{19}$.
}
\end{figure}

The local average order parameter $\langle \mbox{\boldmath$\zeta$}\rangle$ in the long-time limit is illustrated in figure \ref{fig: 4} for a \lq journal bearing' flow (that is, a two-dimensional flow between two non-slip non-concentric rotating boundaries), using the visualisation method illustrated in figure \ref{fig: 2}. The details of this example will be discussed in section \ref{sec: 5}, but this figure indicates that in some cases the solution of this problem may be very complicated.

In two recent works \cite[]{Wil+08,Bez+09}, we have considered the alignment of rheoscopic fluids in response to generic, time-dependent flows. If a time-dependent flow does not decay (for example, if the fluid is continuously stirred), then there is a usually a positive Lyapunov exponent (meaning that infinitesimal separations between fluid elements grow exponentially). In the case of random flows we also find singularities which are related to those in fingerprints. However, there is an important distinction. In the case we consider here the fingerprint-like singularities only emerge after performing a local average of the order parameter in the long-time limit. In a random flow, by contrast, singularities may be observed at short times. The reasons for the difference are explained at the end of this paper.

We assume for simplicity that the crystals are axisymmetric. Our discussion applies in the case where the bodies are much smaller than any length scale of the flow, and for this reason we describe them as \lq particles' for the remainder of this paper. For the case of a steady two-dimensional flow, the limit of infinite aspect ratio is a singular case, and we retain the aspect ratio $\beta$ of the particles as a parameter in our equations. The equation of motion for the unit vector ${\bf n}$ aligned with the symmetry axis of the microscopic axisymmetric particles was originally obtained by \cite{Jef22}. An elegant solution of this equation of motion was subsequently given by \cite{Sze93}, who showed how the solution of this non-linear equation may be obtained by normalising a vector which evolves according to a companion linear equation (Szeri gives credit to earlier works by \cite{Bre62} and \cite{Lip+88}, but these do not contain the general solution). Most of the other literature has applied concepts from dynamical systems theory to the non-linear system of equations obtained by Jeffery (see, for example \cite[]{Shi+91,Sze+91,Mal+91,Sze+94,Shi+97,Gau+98}), and Szeri's own paper makes very limited use of his general solution.

In this work we combine Szeri's solution with the insight that comes from the analogy between the companion linear equation and the time-independent Schr\"odinger equation in one dimension. Typically, the contours of the stream function are closed curves, so that the trajectory of a fluid element is periodic in time, with a period $T$ which depends upon the contour. The evolution of the companion linear equation for a trajectory on a closed contour is analogous to the propagation of the solution of a time-independent Schr\"odinger equation in a spatially periodic potential \cite[]{Zim76}. The solution of this latter problem has bands of energy (the {\em Bloch bands}) where generalised eigenstates exist (which take the form of {\em Bloch waves}), interspersed by {\em band gaps}, intervals of energy for which the electron cannot propagate. The regions where the particles approach a constant direction correspond to the band gaps in the solution of the Schr\"odinger equation, and the Bloch bands correspond to the regions where the particles tumble. In the following we refer (for reasons which will be discussed in section \ref{sec: 5}) to the regions where particles align as {\em hyperbolic bands}, and the regions where they tumble as {\em elliptic bands}. The analogy with solid-state physics will be useful to readers who are familiar with that field, but it is not essential to understanding the paper.

In section \ref{sec: 2} we discuss the equation of motion for the direction vector of a single particle. We describe its general solution, and also consider the instructive special case of flows with a uniform velocity gradient. In section \ref{sec: 3} we consider the order parameter, explaining the motivation for the definition which was used by \cite{Bez+09}, and explaining how the order parameter is related to the reflection of light from the system. Section \ref{sec: 4} discusses the case of recirculating flows in two dimensions, including an analogy with Bloch's theorem and ideas related to Anderson localisation (described by \cite{Zim76,Mot+61}). We show that the order parameter has a quasiperiodic structure in regions where the rods tumble. Section \ref{sec: 5} considers the calculation of the locally averaged order parameter in the long-time limit, and considers topological aspects of the solution. We show that the Poincar\'e index can change by at most $\pm\frac{1}{2}$ on crossing a band (with simple annular topology) where the rods tumble, and give a criterion for determining the change of Poincar\'e index. If the Poincar\'e index changes upon crossing an elliptic band, there must be a singularity of the average order parameter. In section \ref{sec: 5} we identify normal forms for these singularities. Section \ref{sec: 6} contains some concluding remarks, and a discussion of how the results of this paper differ from the case of random flows, considered by \cite{Wil+08,Bez+09}.


\section{Equation of motion and its solution}
\label{sec: 2}

We consider the motion of very small rigid axisymmetric particles immersed in a fluid flow with velocity field $\mbox{\boldmath$v$}(\mbox{\boldmath$r$},t)$. We assume they are small compared to the characteristic length scale of the flow, and also sufficiently small that they do not interact with each other or perturb the velocity field. They are assumed to have negligible inertia so that their motion is dominated by viscous forces. The equation of motion of the centre of the particle $\mbox{\boldmath$r$}(t)$ is then the advective equation $\dot{\mbox{\boldmath$r$}}=\mbox{\boldmath$v$}(\mbox{\boldmath$r$},t)$ (we shall use dots to indicate time derivatives).
For sufficiently small particles, the equation of motion for ${\bf n}(t)$ can only depend upon the gradient of the velocity field, described by a tensor ${\bf A}(t)$ with matrix elements $A_{ij}(t)=\partial v_i/\partial r_j(\mbox{\boldmath$r$}(t),t)$, where $\mbox{\boldmath$r$}(t)$ is the particle trajectory. The motion of a unit vector ${\bf n}$ aligned with the axis of symmetry is determined by the condition that the torque on the particle is equal to zero. \cite{Jef22} determined the equation of motion for ${\bf n}(t)$ in the case of a spheroidal particle, and \cite{Bre62} showed that the equation of motion for a general axisymmetric particle is of the same form. The equation of motion may be written as
\begin{equation}
\label{eq: 2.1}
\frac{{\rm d}{\bf n}}{{\rm d}t} = {\bf B} {\bf n} -{\bf n} ({\bf n}\cdot {\bf B}{\bf n})
\end{equation}
where ${\bf B}$ is a matrix derived from the rate of strain matrix ${\bf A}$ as follows
\begin{equation}
\label{eq: 2.2}
{\bf B} = \alpha_1 {\bf A} -\alpha_2 {\bf A}^{\rm T}\ ,\ \ \alpha_1+\alpha_2=1.
\end{equation}
Here ${\bf A}^{\rm T}$ is the transpose of the matrix ${\bf A}$, and $\alpha_1$, $\alpha_2$ are dimensionless parameters which are determined by the aspect ratio of the particle: for an ellipsoid of aspect ratio $\beta $ (with $\beta\ge 1$), Jeffery showed that $\alpha_1=\beta^2/(\beta^2+1)$, $\alpha_2=1/(\beta^2+1)$. Jeffery originally wrote his equations of motion in component form, however our (\ref{eq: 2.1}), (\ref{eq: 2.2}) are equivalent to equations (10) and (12) in \cite{Mal+91} with ${\bf E} = {1\over 2}({\bf A}+{\bf A}^{\rm T})$ and $\mbox{\boldmath$\omega$} = \mbox{\boldmath$\nabla$}\wedge \mbox{\boldmath$u$}$.

\cite{Jef22} also discussed the particular case of the motion of an ellipsoid of revolution in a uniform shear flow, and showed that (except for the limiting case of a rod) the particle exhibits a tumbling motion, which has been observed experimentally by \cite{Sav85}. This has been extended to consider the tumbling motion of particles in cellular flows \cite[]{Mal+91}. The case of a general linear flow was discussed by \cite{Sze+91}, who also discussed the response to a more general flow field in the language of dynamical systems theory. Most other works (for example, \cite{Shi+91,Sze+94,Shi+97,Gau+98}) have also used a dynamical systems approach based upon the nonlinear equation of motion for ${\bf n}$.

We can, however, solve (\ref{eq: 2.1}) in terms of the solution of an auxiliary problem, which is linear. Specifically, we solve the equation
\begin{equation}
\label{eq: 2.3}
\frac{{\rm d}\mbox{\boldmath$d$}}{{\rm d}t}={\bf B}(t)\mbox{\boldmath$d$}
\end{equation}
to determine a vector $\mbox{\boldmath$d$}(t)$. Here  ${\bf B}(t)$ is the matrix ${\bf B}$ evaluated at the position reached by the particle at time $t$, that is ${\bf B}(t)=\alpha_1{\bf A}(\mbox{\boldmath$r$}(t),t)-\alpha_2{\bf A}^{\rm T}(\mbox{\boldmath$r$}(t),t)$. This equation is solved with the initial condition $\mbox{\boldmath$d$}(t_0)={\bf n}_0$, where ${\bf n}_0$ is the initial orientation of the particle at time $t_0$. Now multiply $\mbox{\boldmath$d$}(t)$ by a scalar $\mu(t)$, chosen such that ${\bf n}(t)=\mu (t)\mbox{\boldmath$d$}(t)$ is a unit vector. We find that this normalised vector does indeed satisfy equation (\ref{eq: 2.1}). We therefore have a solution of the nonlinear equation for ${\bf n}(t)$ in the form
\begin{equation}
\label{eq: 2.4}
{\bf n}(t)=\frac{\mbox{\boldmath$d$}(t)}{\vert\mbox{\boldmath$d$}(t)\vert}
\ .
\end{equation}
This is an exact and completely general solution for the orientation, in terms of the solution of a companion linear problem, equation (\ref{eq: 2.3}). Because of the superposition principle, it is almost always much easier to analyse a linear problem, even in circumstances where exact solutions are not available. We exploit this advantage in the remainder of this paper.
This solution was first obtained by \cite{Sze93}, but remarkably most subsequent papers did not make use of this powerful result.

Solving equation (\ref{eq: 2.4}) is sufficient for determining the motion of a single particle with a specified initial orientation, but in many cases we wish to consider the motion of many small particles, or to obtain the solution for an arbitrary initial orientation. In this more general context, instead of solving (\ref{eq: 2.4}) we determine a matrix ${\bf M}(t)$ which is the solution of
\begin{equation}
\label{eq: 2.5}
\frac{\rm d}{{\rm d}t}{\bf M}={\bf B}(\mbox{\boldmath$r$}(t),t)\,{\bf M}
\end{equation}
with initial condition ${\bf M}(0)={\bf I}$ (the identity matrix). Given this matrix, the solution of (\ref{eq: 2.5}) is $\mbox{\boldmath$d$}(t)={\bf M}(t)\mbox{\boldmath$d$}_0$, for any choice of $\mbox{\boldmath$d$}_0$, so that a single solution suffices for all initial directions. A further generalisation is to consider an arbitrary initial position for the particle at time $t_0$. Let ${\bf M}(\mbox{\boldmath$r$},t,t_0)$ be the solution of (\ref{eq: 2.1}) for a particle which reaches position $\mbox{\boldmath$r$}$ at time $t$, having started at $\mbox{\boldmath$r$}_0$ at time $t_0$. In this most general case the orientation is a vector field, ${\bf n}(\mbox{\boldmath$r$},t)$, and our exact solution becomes:
\begin{equation}
\label{eq: 2.6}
{\bf n}(\mbox{\boldmath$r$},t)=
\frac{{\bf M}(\mbox{\boldmath$r$},t,t_0){\bf n}(\mbox{\boldmath$r$}_0,t_0)}
{\vert{\bf M}(\mbox{\boldmath$r$},t,t_0){\bf n}(\mbox{\boldmath$r$}_0,t_0)\vert}\ .
\end{equation}

In the case of rod-like particles, where $\beta\to \infty$, the matrix ${\bf B}(t)$ is equal to the velocity-gradient matrix ${\bf A}(t)$, with elements $A_{ij}=\partial v_i/\partial r_j$. In this case the matrix ${\bf M}(t)$ has a simple physical interpretation, and in the following we use ${\bf M}_A(t)$ to denote the solution of (\ref{eq: 2.3}) in the special case where ${\bf B}={\bf A}$. Consider the trajectories of two particles advected with the fluid: a reference particle with trajectory $\mbox{\boldmath$r$}(t)$, and a nearby particle with trajectory $\mbox{\boldmath$r$}(t)+\delta \mbox{\boldmath$r$}(t)$. To leading order in the separation $\vert \delta \mbox{\boldmath$r$}\vert$, the separation vector is determined by the matrix ${\bf M}_A(t)$: we have
$\delta \mbox{\boldmath$r$}(t)={\bf M}_A(t)\,\delta \mbox{\boldmath$r$}(0)$.
In the language of dynamical systems theory, a matrix with this property is termed a {\em monodromy} matrix. We consider volume preserving flows, so that ${\rm tr}[{\bf A}(t)]=0$. From (\ref{eq: 2.2}), the matrix ${\bf B}(t)$ also has the property that ${\rm tr}[{\bf B}(t)]=0$, as if it were the velocity-gradient of some ficticious volume-preserving flow, and consequently ${\rm det}[{\bf M}(t)]=1$. We will therefore refer to the solution ${\bf M}(t)$ of (\ref{eq: 2.5}) as the {\em pseudomonodromy} matrix of the flow. For the case of rod-like particles, where $\alpha_1=1$ and $\alpha_2=0$ in (\ref{eq: 2.2}), it is the same as the true monodromy matrix of the flow.

The degree to which the solution can be presented in closed form depends upon the specifics of the flow field. First we comment on the exactly solvable case of a time-independent flow with constant velocity gradient ${\bf A}$, because this case already exhibits solutions showing both alignment and tumbling. In this case the matrix ${\bf B}$ is also a constant, and the solution of the linear auxiliary equation (\ref{eq: 2.5}) is $\mbox{\boldmath$d$}(t)={\bf M}(t)\mbox{\boldmath$d$}_0=\exp({\bf B}t)\mbox{\boldmath$d$}_0$. The matrix $\exp({\bf B}t)$ may be expressed in terms of the eigenvalues and eigenvectors of ${\bf B}$. The matrix ${\bf B}$ is typically non-Hermitian, and correspondingly its eigenvectors need not be orthogonal. The behaviour of the solution is determined by the eigenvalues $\lambda_i$. We consider incompressible flow, implying that ${\rm tr}[{\bf B}]=0$, so the eigenvalues sum to zero. We describe both the three-dimensional and two-dimensional cases below (the following discussion overlaps some comments made by \cite{Sze93}).

Apart from degenerate cases, the spectrum may take one of three forms in three dimensions:

\begin{enumerate}

\item Eigenvalues real and distinct, with at least one of them positive. The axis of the particle aligns with the eigenvector $\mbox{\boldmath$u$}_+$ corresponding to the largest eigenvalue, $\lambda_+$ of ${\bf B}$ (in the following we refer to these as the {\em dominant} eigenvector and eigenvalue).

\item There may be a real and positive eigenvalue $\lambda_+$, and a complex pair with negative real part. In this case the axis also aligns with the dominant eigenvector, $\mbox{\boldmath$u$}_+$.

\item There may be two complex conjugate eigenvalues with positive real part (so that the real eigenvalue is negative), with complex conjugate eigenvectors. When these two eigenvectors are combined with complex-conjugate coefficients, the resulting real vector lies in a plane. In the long-time limit the vector $\mbox{\boldmath$d$}(t)$ spirals outwards in this plane. This case corresponds to a tumbling motion of the particle.

\end{enumerate}

Another way to understand the dynamics of the vector $\mbox{\boldmath$d$}(t)$ is to write ${\bf M}(t)$ as a normal form:
\begin{equation}
\label{eq: 2.7}
{\bf M}(t)={\bf X}\,{\bf N}(t)\,{\bf X}^{-1}
\end{equation}
where ${\bf X}$ and ${\bf N}(t)$ are real-valued matrices. In case 1, the matrix ${\bf N}(t)$ is diagonal, with diagonal entries $\exp(\lambda_i t)$. In cases 2 and 3, the matrix ${\bf N}(t)$ is in block-diagonal form with a $2\times 2$ block describing a spiralling motion,
\begin{equation}
\label{eq: 2.8}
{\bf N}(t)=\left(\begin{array}{ccc}
\exp(-\frac{1}{2}\lambda t)\cos(\omega t) &\exp(-\frac{1}{2}\lambda t)\sin(\omega t)&0\cr
-\exp(-\frac{1}{2}\lambda t)\sin(\omega t)&\exp(-\frac{1}{2}\lambda t)\cos(\omega t)&0\cr
0                             &          0                  &\exp(\lambda t)
\end{array}\right)\ .
\end{equation}
Here $\lambda$ is the real eigenvalue of ${\bf B}$, and the complex eigenvalues are $-\frac{1}{2}\lambda \pm {\rm i}\omega$. The spiralling motion may be attractive (spiralling-in, when $\lambda>0$), which is case 2, or repelling (spiralling-out, when $\lambda<0$), which is case 3.

In the case of two-dimensional incompressible flow, the matrix ${\bf M}$ may have two reciprocal real eigenvalues (the {\em hyperbolic} case), or else two complex conjugate eigenvalues which lie on the unit circle (the {\em elliptic} case).
In the hyperbolic case the vector $\mbox{\boldmath$d$}(t)$ comes into alignment with the eigenvector of ${\bf B}$ which corresponds to the positive eigenvalue (the dominant eigenvector). In the elliptic case, where ${\bf B}$ has purely imaginary eigenvalues $\pm{\rm i}\omega$, the pseudomonodromy matrix ${\bf M}(t)$ can be expressed in terms of a normal form, analogous to (\ref{eq: 2.8}), with ${\bf N}(t)$ replaced by a rotation matrix ${\bf R}(\omega t)$ representing rotation in the plane by an angle $\omega t$.

The matrix ${\bf B}$ is typically non-Hermitian, and correspondingly its eigenvectors need not be orthogonal. We briefly consider the consequences of this observation in the two-dimensional case. If the matrix ${\bf B}$ is elliptic, the vector $\mbox{\boldmath$d$}(t)$ rotates, the linear transformation ${\bf X}$ in (\ref{eq: 2.7}) transforms the circular motion of ${\bf R}(\omega t){\bf X}^{-1}\mbox{\boldmath$d$}_0$ to motion on an ellipse. In some circumstances this ellipse may have a large aspect ratio. In this case the vector ${\bf n}(t)=\mbox{\boldmath$d$}(t)/\vert\mbox{\boldmath$d$}(t)\vert$ will spend most of its time nearly aligned with the long axis of the ellipse, reversing direction rapidly at times separated by $\pi/\omega$.


\section{Order parameter and light scattering}
\label{sec: 3}

\subsection{General definition of the order parameter}
\label{sec: 3.1}

The alignment of the particles may be described by an order parameter. If the rheoscopic fluid is left to stand for a while, the crystals become randomly oriented due to Brownian motion. When the fluid is set in motion, the crystals start to align and at later times we can describe the distribution of angles by a probability density. In the case we consider below the particles are aligned in a plane so their direction is defined by a single angle $\theta$. Because the direction vector is non-oriented, the probability density $P(\theta)$ satisfies $P(\theta+\pi)=P(\theta)$. This probability density will depend upon both position and time, but we suppress the arguments $\mbox{\boldmath$r$}$ and $t$ in the discussion below.

A suitable order parameter for the rod-like particles can be obtained from $P(\theta)$ by first calculating the inertia tensor of the rods, which has components:
\begin{equation}
\label{eq: 3.1}
I_{ij}=\int_0^{2\pi} {\rm d}\theta\ P(\theta) ({\bf n}_i\cdot {\bf n}(\theta))({\bf n}_j\cdot{\bf n}(\theta))
\end{equation}
where ${\bf n}(\theta)$ is a unit vector in the direction $\theta$. The three distinct components of $I_{11}$, $I_{12}$, $I_{22}$ are not independent, because the vector ${\bf n}(\theta)$ is constrained to have unit length. They can be mapped to the order parameter vector $\mbox{\boldmath$\zeta$}$ as  follows. The inertia tensor has real, positive eigenvalues ${\cal I}_1$, ${\cal I}_2$ and corresponding orthonormal eigenvectors $\mbox{\boldmath$U$}_1$, $\mbox{\boldmath$U$}_2$, with ${\cal I}_1\ge {\cal I}_2$. The eigenvalues satisfy ${\cal I}_1+{\cal I}_2=1$, and the case ${\cal I}_1=1$ corresponds to perfect alignment, whereas ${\cal I}_1={\cal I}_2=\frac{1}{2}$ corresponds to an isotropic distribution. We define $\mbox{\boldmath$\zeta$}$ to be a non-oriented vector in the direction $\mbox{\boldmath$U$}_1$ with magnitude which is a function of ${\cal I}_1-{\cal I}_2$. Let us consider a special case where the rods align with the direction $\bar \theta$ with probability $p$, or else are randomly distributed with probability $1-p$, that is
\begin{equation}
\label{eq: 3.2}
P(\theta)=\frac{p}{2}[\delta(\theta-\bar\theta)+\delta(\theta-\bar\theta-\pi)]+\frac{1-p}{2\pi}
\ .
\end{equation}
It is natural to define the order parameter so that $\mbox{\boldmath$\zeta$}=p{\bf n}(\bar\theta)$ in this case. For this distribution, in the case $\bar\theta=0$ we find ${\cal I}_1=(1+p)/2$ and ${\cal I}_2=(1-p)/2$, so that ${\cal I}_1-{\cal I}_2=p$. We therefore define the order parameter as
\begin{equation}
\label{eq: 3.3}
\mbox{\boldmath$\zeta$}=({\cal I}_1-{\cal I}_2)\mbox{\boldmath$U$}_1
\ .
\end{equation}
This is a general definition for the order parameter of rod-like particles in two dimensions. An analogous definition can be used in three dimensions, where a general inertia tensor has six independent components, but the inertia tensor for the rod directions has five parameters because of the constraint that $\vert{\bf n}\vert=1$.

\subsection{Order parameter in terms of the monodromy matrix}
\label{sec: 3.2}

Let us consider the evaluation of this order parameter for the case where the particles are initially randomly oriented, so that the initial direction ${\bf n}_0$ in (\ref{eq: 2.2}) is uniformly distributed about the unit circle. According to the solution presented in section \ref{sec: 2}, a vector ${\bf n}_0$ on this circle is mapped to a vector $\mbox{\boldmath$d$}(t)$ which lies on an ellipse. This ellipse is described by its aspect ratio, $\nu \ge 1$, and by the direction of its longest axis, $\bar\theta$. In the following we obtain the probability density $P(\theta)$ and use this to obtain the order parameter $\mbox{\boldmath$\zeta$}$ in terms of $\nu$ and $\bar\theta$.

\begin{figure}
\centerline{\includegraphics[width=5cm]{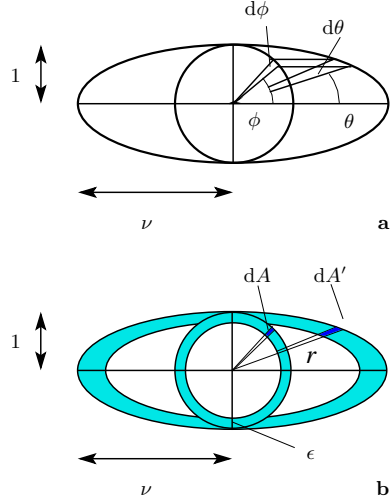}}
\caption{\label{fig: 5} Illustrating the geometrical construction used to determine the probability density for the angle, $P(\theta)$.
}
\end{figure}

An angle interval ${\rm d}\phi$ on the unit circle is mapped to a segment of the ellipse which is at an angle $\theta$ to its longer axis, and which spans an angle interval ${\rm d}\theta$. The angle $\theta$ is independent of the overall scale of the ellipse, and we find it convenient to consider the case where the short axis intersects the unit circle (see figure \ref{fig: 5}{\bf a}, where $\bar\theta=0$ so that the long axis is horizontal). The probability element for the direction of ${\bf n}_0$ being in the original interval is ${\rm d}P={\rm d}\phi/2\pi$. This is the same as the probability element for $\mbox{\boldmath$d$}(t)$ being in the interval ${\rm d}\theta$ on the ellipse, so that the probability density $P(\theta)$ satisfies
\begin{equation}
\label{eq: 3.4}
{\rm d}P=\frac{1}{2\pi}{\rm d}\phi=P(\theta){\rm d}\theta
\ .
\end{equation}

An elementary geometrical construction can be used to surmise the relation between ${\rm d}\phi$ and ${\rm d}\theta$. Instead of considering the mapping of a circle to an ellipse, let us consider the image of a narrow annulus of angular width ${\rm d}\phi$ between a circle with unit radius and one with radius $1-\epsilon$ (with $\epsilon\ll 1$), so that the area of this element is ${\rm d}A\sim \epsilon {\rm d}\phi$. The element of the annulus is the set difference between two segments of discs spanned by an angle ${\rm d}\phi$, one of unit radius, the other of radius $1-\epsilon$. These segments are transformed into regions which may also be approximated by segments of circles: the larger one is approximated by a segment of a circle radius $r$ spanned by an angle ${\rm d}\theta$, having area $\sim \frac{1}{2}r^2{\rm d}\theta$ and the smaller one by a segment which is smaller in area by a factor $(1-\epsilon)^2\sim 1-2\epsilon$ (see figure \ref{fig: 5}{\bf b}). The area of the transformed image of the annulus is therefore ${\rm d}A'=\epsilon r^2{\rm d}\theta$. Because the transformation from a circular region to an ellipse stretches the $x$-axis by the factor $\nu$, we also have ${\rm d}A'=\nu \epsilon {\rm d}\phi$. We conclude that ${\rm d}\phi=r^2{\rm d}\theta/\nu$, where $r$ is the distance from the origin to a point on the ellipse at angle $\theta$ from the long axis. The equation of the ellipse is $\nu^2=x^2+\nu^2 y^2$, where $x=r\cos\theta$, $y=r\sin\theta$, so that $\nu^2=r^2[(\nu^2-1)\sin^2\theta+1]$. Using (\ref{eq: 3.4}) we therefore conclude that the probability density for the direction of the vector $\mbox{\boldmath$d$}$ in (\ref{eq: 2.4}) is
\begin{equation}
\label{eq: 3.5}
P(\theta)=\frac{r^2}{2\pi \nu}=\frac{\nu}{2\pi}\frac{1}{(\nu^2-1)\sin^2(\theta-\bar\theta)+1}\ .
\end{equation}
Using the identities
\begin{eqnarray}
\label{eq: 3.6}
\int_0^{2\pi}{\rm d}x \ \frac{\cos^2x}{A\sin^2 x+1}&=&2\pi\frac{\sqrt{A+1}-1}{A}
\nonumber \\
\int_0^{2\pi}{\rm d}x \ \frac{\sin^2x}{A\sin^2 x+1}&=&2\pi\frac{\sqrt{A+1}-1}{A\sqrt{A+1}}
\end{eqnarray}
we find that for this probability density the elements of the inertia tensor are
\begin{eqnarray}
\label{eq: 3.7}
I_{11}=1-I_{22}&=&\frac{\nu}{\nu +1}\cos^2\bar\theta+\frac{1}{\nu+1}\sin^2\bar\theta
\nonumber \\
I_{12}&=&\frac{\nu-1}{\nu+1}\cos\bar\theta\sin\bar\theta
\ .
\end{eqnarray}
The eigenvalues of the inertia tensor are then ${\cal I}_1=\frac{\nu}{\nu+1}$ and ${\cal I}_2=\frac{1}{\nu+1}$.
The order parameter for an initially uniform angular distribution is therefore
\begin{equation}
\label{eq: 3.8}
\mbox{\boldmath$\zeta$}=\frac{\nu-1}{\nu+1} {\bf n}(\bar\theta)
\end{equation}
where ${\bf n}(\theta)$ is a unit vector in the direction $\theta$. It remains to express the aspect ratio $\nu\ge 1$ of the ellipse in terms of the matrix ${\bf M}$. The equation defining the unit circle $\vert{\bf n}_0\vert=1$ can be written $\mbox{\boldmath$x$}\cdot\mbox{\boldmath$x$}=1$. In terms of $\mbox{\boldmath$x$}'={\bf M}\mbox{\boldmath$x$}$, this condition becomes the equation for an ellipse: $\mbox{\boldmath$x$}'\cdot {\bf K}\mbox{\boldmath$x$}'=1$, with
\begin{equation}
\label{eq: 3.9}
{\bf K}=({\bf M}^{-1})^{\rm T}{\bf M}^{-1}
=({\bf M}{\bf M}^{\rm T})^{-1}\ .
\end{equation}
The aspect ratio $\nu$ is therefore the square root of the ratio of the eigenvalues of the real, symmetric positive definite matrix ${\bf K}$. This may also be determined from the ratio of the eigenvalues of ${\bf K}^{-1}={\bf M}{\bf M}^{\rm T}$. If the matrix ${\bf K}^{-1}$ has eigenvalues $\lambda_1$, $\lambda_2$ with corresponding orthonormal eigenvectors $\mbox{\boldmath$U$}_1$, $\mbox{\boldmath$U$}_2$ ordered so that $\lambda_1>\lambda_2$, then the parameters in (\ref{eq: 3.8}) are then $\nu=\sqrt{\lambda_1/\lambda_2}$ and ${\bf n}(\bar \theta)=\mbox{\boldmath$U$}_1$.

In a generic flow, the matrix ${\bf B}(t)$ is neither constant nor periodic, and we expect that the solution of (\ref{eq: 2.3}) will have a positive largest Lyapunov exponent. In this case, the pseudomonodromy matrix ${\bf M}$ will become hyperbolic almost everywhere, having a unique largest eigenvalue, which increases as time increases. The rods will then align very close to the direction of the dominant eigenvector, irrespective of their initial orientation. However, if the pseudomonodromy matrix remains elliptic, there is no dominant eigenvector and the final direction remains dependent upon the initial orientation. In the hyperbolic case where the rods approach perfect alignment, the order parameter vector approaches a unit vector, but in the elliptic case it is shorter than unit length.

\subsection{Relating the order parameter to light scattering}
\label{sec: 3.3}

The order parameter can be investigated experimentally by examining the reflection of light by the rheoscopic fluid. Because we are primarily interested in two-dimensional flows, we consider how the light scattering may be related to the order parameter in the case where the illumination is confined to a surface. By way of examples, this is relevant when the rheoscopic fluid is a thin layer floating on a denser, immiscible fluid, or when the rheoscopic agent is used without dilution, so that the optical depth is very small (implying that scattered light comes from a thin layer close to the surface). The image contrast is greatest when the illumination comes from a direction in the same plane as the surface, and we choose to specify its direction by means of the angle $\phi$ of the direction perpendicular to that from which the beam is incident.

The intensity of light reflected by the microscopic crystals depends upon their orientation relative to the direction of the source of the light. The angular dependence of the scattering depends upon a variety of factors, of which the ratio of the size of the crystals to the wavelength of light and their surface roughness are important. If the crystals are aligned with their long axis at angle $\theta$, the intensity of the scattered light will be $f(\phi-\theta)$, for some function $f$ which is even and periodic with period $\pi$. In the following we consider the limit where the crystals are smaller than the wavelength of light, in which case the amplitude of the scattered radiation is proportional to the projected area of the crystal in the direction of the incident light. This implies that a rod at angle $\theta$ scatters light from a source which is perpendicular to the direction $\phi$ with an intensity proportional to $\cos^2(\theta-\phi)+\gamma$, where $\gamma$ is a contribution arising from diffuse background scattering. In our subsequent discussion we shall use this form for the scattering kernel, with $\gamma=0$.

More detailed information about the orientation of the particles may be revealed by using three different light sources with different colours, illuminating the fluid from three different directions. The intensity of the scattering of light from a given source depends upon the direction of the particle relative to the direction of the light source. At any given position the fluid reflects with a colour $C$ determined by additive mixing of the scattered light from red, green and blue ($R$, $G$, $B$) sources, which we assume are arranged about the sample at directions separated by $120^\circ$, as illustrated in figure \ref{fig: 2}. This results in the light being scattered with a colour $C$ which is determined by additive mixing of the primary colours $R$, $G$, $B$:
\begin{equation}
\label{eq: 3.10}
C=I(0)\,R+I(2\pi/3)\,G+I(4\pi/3)\,B
\end{equation}
where in the limiting case of short rods $I(\theta)$ is the inertia of the axial distribution relative to the direction $\theta$:
\begin{equation}
\label{eq: 3.11}
I(\theta)=\int_0^{2\pi}{\rm d}\theta'\ P(\theta')\cos^2(\theta-\theta')\ .
\end{equation}
In principle just two of the functions $I(0)$, $I(2\pi/2)$ and $I(4\pi/3)$ are sufficient to determine the two parameters of the order parameter. However, using three colours has two advantages: with three colours the ratios of the scattered intensities can be used, so that the normalisation of the intensities is not relevant. Also, as shown by \cite{Bez+09}, with three colours the Poincar\'e index of singularities can be visualised directly. The mapping between the order parameter $\mbox{\boldmath$\zeta$}$ and the colour $C$ of the reflected light is illustrated in figure \ref{fig: 2}.
The use of coloured light sources to enhance rheoscopic images was previously suggested by \cite{Tho+99}. Their work does not consider the relation between the colour mixing and the ordering of the particles.

For larger crystals the function $\cos^2(\theta-\theta')$ is replaced by another function $f(\theta-\theta')$. This would make a quantitative but not a qualitative difference to the colour images which are displayed here.


\section{Steady flows in two dimensions}
\label{sec: 4}

\subsection{Hyperbolic and elliptic bands}
\label{sec: 4.1}

Now we turn to considering rheoscopic particles in a steady incompressible flow where the velocity vector is confined to a plane with coordinates $(x,y)$. This system was previously considered by \cite{Sze93}, who showed that there are regions (which we term {\em hyperbolic bands}) where the particles align with each other. His paper gives a treatment of equation (\ref{eq: 2.1}) using concepts from dynamical systems theory. In particular, the regions in which the particles align are determined by looking for stable fixed points of the Poincar\'e map for the particle axis direction as it is advected around a contour of the stream function. Below we use the solution (\ref{eq: 2.6}), which gives a more thorough insight into this system, as well as being more computationally efficient (because it is not necessary to repeat the calculation for different initial directions of the rod). Together with the quadratic form for determining the order parameter, (\ref{eq: 3.9}), this also allows us to describe the alignment of the particles in the regions outside the hyperbolic bands.

The velocity field of a steady, incompressible two-dimensional flow may be derived from a stream function $\psi(x,y)$: we have $\mbox{\boldmath$v$}=(\partial \psi/\partial y,-\partial \psi/\partial x)$. By analogy with Hamiltonian's equations of motion for a one-freedom autonomous system, we see that the trajectories follow contours of the stream function, so that a trajectory labelled by $\psi_0$ is defined by writing $\psi(x,y)=\psi_0$. The contours may be either closed or open. Particles which are advected along a closed contour have a periodic motion, with a period $T$ (which is a function of $\psi_0$). This periodicity simplifies the analysis of the behaviour of advected particles, and we concentrate on the periodic case. (Periodic behaviour can also occur if $\psi (x,y)$ is periodic in one or both variables, and our discussion is readily extended to such cases).

We have seen that the behaviour of axisymmetric particles is determined by the pseudomonodromy matrix ${\bf M}(t)$. In the two-dimensional incompressible case this matrix is a $2\times 2$ matrix which satisfies ${\rm det}[{\bf M}(t)]=1$. Such a matrix is either hyperbolic, having two reciprocal real eigenvalues, or elliptic, with two mutually conjugate complex eigenvalues with modulus equal to one. The character of this matrix is readily determined from its trace: if $\vert {\rm tr}[{\bf M}]\vert > 2$, the matrix is hyperbolic, whereas if $\vert {\rm tr}[{\bf M}]\vert<2$, the matrix is elliptic (and if $\vert {\rm tr}[{\bf M}]\vert=2$, the matrix is a shear). By comparison with the case of constant matrix ${\bf B}$ which was discussed in section \ref{sec: 2}, we anticipate that if the matrix ${\bf M}$ is hyperbolic, the advected particles tend to approach a given direction, whereas the elliptic case is associated with tumbling motion. This expectation turns out to be correct, in a qualified sense as discussed below.

We can label points on a closed trajectory by the time $t_0$ taken to reach the point from an arbitrary reference point on the orbit. Let ${\bf M}(t,t_0)$ be the pseudomonodromy matrix for the trajectory which starts at time $t_0$ and at the point labelled by $t_0$, ending a time $t$. Let us consider the evaluation of ${\bf M}(t,t_0)$, in the case where $t$ is written in the form $t=t_1+NT$ (where $T$ is the period and $N$ an integer). We can express this general matrix in terms of a pseudomonodromy matrix for a single cycle, ${\bf M}_0={\bf M}(T,0)$, together with matrices representing short time evolution for a fraction of a cycle. We can write
\begin{equation}
\label{eq: 4.1}
{\bf M}(t,t_0)={\bf M}(t_1,0)[{\bf M}_0]^N{\bf M}^{-1}(t_0,0)
\ .
\end{equation}
This shows that the long-time behaviour is determined by the character of the matrix ${\bf M}_0$, which can be computed by propagating a solution of (\ref{eq: 2.5}) for a finite time. In particular, if ${\bf M}_0$ is hyperbolic, the matrix ${\bf M}(t,t_0)$ will have one eigenvalue which is much larger than the other when $t-t_0\to \infty$. Because the eigenvalues of a matrix are invariant under a similarity transform, the structure of (\ref{eq: 4.1}) implies that the character of ${\bf M}_0$ (hyperbolic or elliptic) is independent of the choice of starting point on the contour.

Because the elliptic or hyperbolic character of a trajectory is independent of its starting point, we can label the contours of the stream function according to the character of the one-period monodromy matrix, ${\bf M}_0$. From (\ref{eq: 4.1}), we see that when $t_1=0$ and when ${\bf M}_0$ is hyperbolic, the particles align with the eigenvector of ${\bf M}_0$ corresponding to its largest eigenvector. More generally, in the hyperbolic case the orientation at any position aligns with the direction of the dominant eigenvector of the monodromy matrix for the one-period orbit which ends at that position. On contours where ${\bf M}_0$ is elliptic, at any given position the particles continue to tumble as $t-t_0\to \infty$. The contours of $\psi(x,y)$ are may therefore be divided into {\em elliptic bands}, where the pseudo-monodromy matrix is elliptic and the particles tumble, and {\em hyperbolic bands}, where the pseudo-monodromy matrix is hyperbolic and where the direction approaches a constant vector field. These bands are analogous to the bands which occur for the solution of the Schr\" odinger equation for a one-dimensional potential \cite[]{Zim76}, where the {\em transfer matrix} for solutions of the Schr\"odinger equation plays the same role as the monodromy matrix for a single orbit ${\bf M}_0$. The hyperbolic bands correspond to the {\em band-gaps} in the solution of the Schr\"odinger equation, where the wavefunction increases exponentially in one direction, so that there are no satisfactory eigenstates. The elliptic bands correspond to the {\em energy bands} of the Schr\"odinger equation, where its solutions are Bloch waves \cite[]{Zim76}. We remark that our discussion may also be viewed as an example of the application of Floquet theory. We emphasise that the hyperbolic bands are the same as the aligning regions in \cite{Sze93}.

We remark that in the special case where the particles are rod-like (that is, $\alpha_2\to 0$ in (\ref{eq: 2.2})), the matrix ${\bf M}(t,t_0)$ is the true monodromy matrix. The one-period monodromy matrix for a periodic two-dimensional flow is always a simple shear, and rod-like particles will always align with the contours of the stream function.

The results in section \ref{sec: 2} above show that in a simple shear flow, the particles always tumble rather than coming into alignment (except for the limiting case where the aspect ratio of the rods is infinite). Because a steady two-dimensional flow locally resembles a shear flow, it might therefore be expected that the transfer matrix ${\bf M}_0$ would always be elliptic, because it can be thought of as a product of matrices each of which would individually be generated by an elliptic flow. This need not be the case, however. It is known from studies of Anderson localisation for the one-dimensional Schr\"odinger equation that products of elliptic matrices can be hyperbolic \cite[]{Mot+61}. (In the context of Anderson localisation, this statement is equivalent to the observation that localised states occur for energies where there is no classical potential barrier \cite[]{Zim76}). We therefore conclude that alignment of particles around periodic trajectories is possible, although it might be argued to be un-expected.

\begin{figure}
\centerline{\includegraphics[width=12cm]{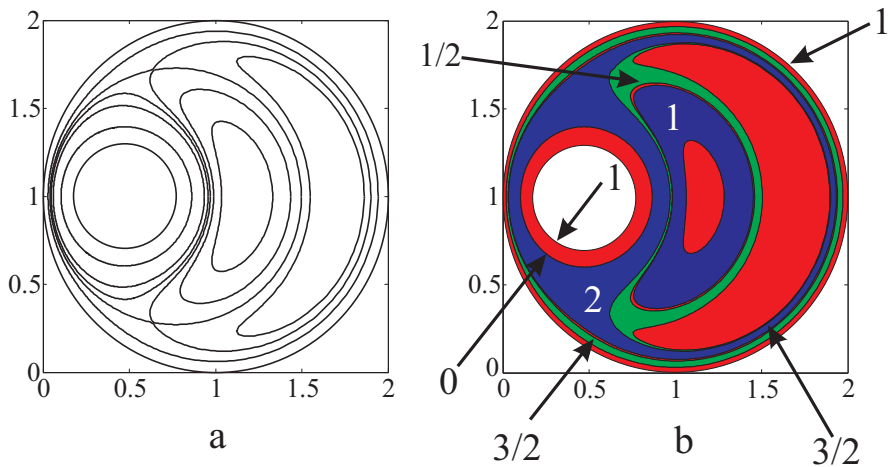}}
\caption{\label{fig: 6} {\bf a} Contours of the stream function $\psi(x,y)$ for a journal bearing system: both circular boundaries rotate in the same direction, with the angular speed of the inner boundary exceeding that of the outer boundary by a factor of $20$. {\bf b} These contours can be coloured according to whether the transfer matrix ${\bf M}_0$ is elliptic $\vert{\rm tr}({\bf M}_0)\vert <2$ (red), or hyperbolic, ${\rm tr}({\bf M}_0)>2$ (blue) and ${\rm tr}({\bf M}_0)<-2$ (green). Each hyperbolic band is labelled by its Poincar\'e index. In this illustration we set $\alpha_1=0.95$, $\alpha_2=0.05$ in equations (\ref{eq: 2.1}), (\ref{eq: 2.2}), (which corresponds to ellipsoidal particles with aspect ratio $\beta=\sqrt{19}=4.36..$).}
\end{figure}

We now turn to consider an example of textures formed by the alignment of axisymmetric particles in steady two-dimensional flows. Figure \ref{fig: 6}{\bf a} displays the contours of the stream function for a two-dimensional flow, exhibiting saddle points as well as extrema. The flow is a \lq journal bearing' flow, where the circular boundaries rotate with different angular velocities. The stream function for this flow was obtained by \cite{Jef22a}, \cite{Mul42}, \cite{Wan50}, and is discussed in detail in \cite[]{Bal+76}. In this example, the walls rotate in the clockwise sense, with the angular velocity of the inner wall exceeding that of the outer wall by a factor of $20$. The radius of the inner wall is $0.3$ times that of the outer wall, and the eccentricity parameter $\bar \varepsilon$ of \cite[]{Bal+76} is $\frac{3}{4}$, so that the centre of the inner boundary is offset by a multiple of $0.525..$ times the radius of the outer wall. This system can be realised physically by filling the space between two vertical rotating cylinders with a rheoscopic fluid, and figure \ref{fig: 4} (which will be explained fully in section \ref{sec: 5}) is an illustration of the complexity of the pattern of light scattering from the surface which could be observed in the long-time limit. The hyperbolic bands (blue or green) and elliptic bands (red) are illustrated in figure \ref{fig: 6}{\bf b}, with the hyperbolic bands shaded blue if ${\rm tr}({\bf M}_0)>2$, green if ${\rm tr}({\bf M}_0)<-2$ (the reason for making the distinction between thee two hyperbolic cases will be considered in section \ref{sec: 5.3}). There is a contour which marks a transition from ${\rm tr}{\bf M}_0>2$ to ${\rm tr}{\bf M}_0<2$ without passing through an elliptic zone: this is possible because the contour is a separartix where the topology of the contours changes. This example of a steady two-dimensional flow is the same as was studied by \cite{Sze93}, and the hyperbolic bands in figure \ref{fig: 6}{\bf b} correspond to the aligning regions which were obtained by Szeri. The Poincar\'e indices of the hyperbolic bands are also shown, and these are equal to one-half of the \lq flip numbers' which were discussed in \cite{Sze93}.

\subsection{The order parameter in elliptic bands}
\label{sec: 4.2}

Let us consider the form of the order parameter in the elliptic bands. Provided the period $T$ of an orbit depends upon the stream function $\psi$, a passive scalar function will be wound into an increasingly tight spiral under the action of a two-dimensional steady flow. Its lines of constant density will become closely aligned with the contours of the stream function, with the scalar having an approximately periodic behaviour when traced in a direction perpendicular to the streamlines. Figure \ref{fig: 1} showed an example of the evolution of the order parameter as time increases, showing the development of an increasingly tight spiral pattern. However, we shall see that the behaviour of the order parameter is more complicated than that of a passive scalar, in that its variation in a direction perpendicular to the contours of $\psi(x,y)$ is quasiperiodic rather than periodic.

\begin{figure}
\centerline{\includegraphics[width=6cm]{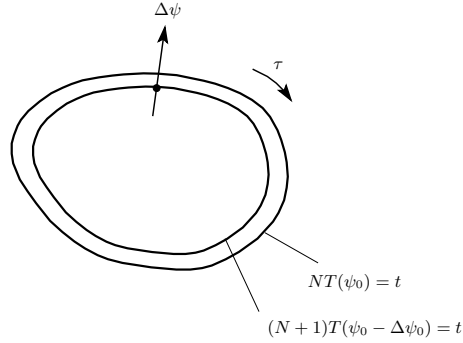}}
\caption{\label{fig: 7} Illustrating the coordinates $\Delta \psi$, $\tau$ which are used in the discussion of elliptic bands.
}
\end{figure}

Consider the variation of the order parameter within an elliptic band as a function of position for large time $t$, in the vicinity of a reference point which lies on a closed contour of $\psi$. In the neighbourhood of this reference point $(x_0,y_0)$ we use two coordinates $\Delta \psi$ and $\tau$ to label points $(x,y)$.  We define $\Delta \psi=\psi(x,y)-\psi(x_0,y_0)$. We define a reference point on other contours of $\psi$ by drawing a line which is perpendicular to the contour passing through $(x_0,y_0)$. We label the distance along a contour by the time $\tau$ taken to reach that point starting from the reference point on the orbit. This coordinate system is illustrated in figure \ref{fig: 7}.

Now let us specialise by taking the reference point to lie on a contour such that $t$ is a multiple of the period $T$, so that $t=NT$ for some integer $N$. For a set of isolated contours the motion will also be periodic, making a different number of orbits in the same time $t$. For large $t$ these contours are approximately evenly spaced, with the spacing $\Delta \psi_0$ of the contours of the stream function being
\begin{equation}
\label{eq: 4.2}
\Delta \psi_0=\left\vert\frac{T}{N}\frac{{\rm d}\psi}{{\rm d}T}\right\vert=\left\vert\frac{1}{N}\frac{{\cal A}'}{{\cal A}''}\right\vert
\end{equation}
where ${\cal A}(\psi)$ is the area enclosed by the contour with stream function $\psi$.

The transfer matrix may be written in terms of its normal form, similar to (\ref{eq: 2.7}). First consider the form of this matrix along the line $\tau=0$. At $(x,y)=(x_0,y_0)$, we have ${\bf M}(0,0)={\bf M}_0^N$, where ${\bf M}_0$ is the transfer matrix (that is, the pseudomonodromy matrix for one orbit). We write the transfer matrix in normal form as follows:
\begin{equation}
\label{eq: 4.3}
{\bf M}_0={\bf X}\,{\bf R}(\theta_0)\,{\bf X}^{-1}
\end{equation}
where ${\bf R}(\theta)$ is a rotation matrix for angle $\theta$. When $\psi$ changes by $\Delta \psi_0$, the trajectory makes one additional orbit, so that the transfer matrix becomes ${\bf M}(\Delta \psi_0,0)={\bf M}_0^{N+1}$. We can therefore write ${\bf M}(\Delta \psi,0)={\bf X}\,{\bf R}(\theta)\,{\bf X}^{-1}\,{\bf Z}(\Delta \psi/\Delta \psi_0)$, where ${\bf Z}(x)$ is a $2\times 2$ matrix which is a periodic function of $x$, with
\begin{equation}
\label{eq: 4.4}
\theta=\theta_0\left(N+\frac{\Delta \psi}{\Delta \psi_0}\right)
\ .
\end{equation}
and
\begin{equation}
\label{eq: 4.5}
{\bf Z}(x+1)={\bf Z}(x)\ , \ \ {\bf Z}(0)={\bf I}
\ .
\end{equation}
With these notations and definitions, for a general position the transfer matrix is
\begin{equation}
\label{eq: 4.6}
{\bf M}(\Delta \psi,\tau)={\bf M}(\tau)\,{\bf X}\,{\bf R}(\theta)\,{\bf X}^{-1}\,{\bf Z}\left(\Delta \psi/\Delta \psi_0\right)\,{\bf M}^{-1}(\tau)
\ .
\end{equation}

In the limit as $N\to\infty$ the order parameter depends increasingly sensitively upon $\psi$, but the sensitivity to $\tau$ is independent of $N$. The dependence of $\mbox{\boldmath$\zeta$}$ upon $\Delta \psi$ is quasiperiodic, being associated with two periods. One period $\Delta \psi_0$ is associated with the change in $\psi$ required for the trajectory to make an additional orbit in time $t$. There is another periodicity associated with the change in $\psi$ required for the phase $\theta$ in (\ref{eq: 4.4}) to increment by $2\pi$. This additional periodicity is $\Delta \psi_1=2\pi \Delta \psi_0/\theta_0$.

Equation (\ref{eq: 4.6}) can be used to obtain a complicated expression for the matrix of the quadratic form describing the order parameter, (\ref{eq: 3.9}), which is arguably too unwieldy to be of much use. However, in the next section we shall see that although the order parameter depends increasingly sensitively on position in the long-time limit, the order parameter of the locally-averaged orientation has a very simple representation.


\section{Averaging, singularities and topology of the order parameter}
\label{sec: 5}

\subsection{Local average of the order parameter}
\label{sec: 5.1}

We have seen that in the elliptic bands the order parameter varies increasingly rapidly as a function of $\psi$ in the limit as $t\to \infty$. Eventually the order parameter fluctuates on a length scale which is small compared to the resolving power of the eye. In this limit it is necessary to perform a local average of the inertia tensor (\ref{eq: 3.1}) representing the distribution of orientations. The order parameter of this locally averaged quantity determines the appearance of the rheoscopic suspension in the long-time limit.

\begin{figure}
\centerline{\includegraphics[width=6cm]{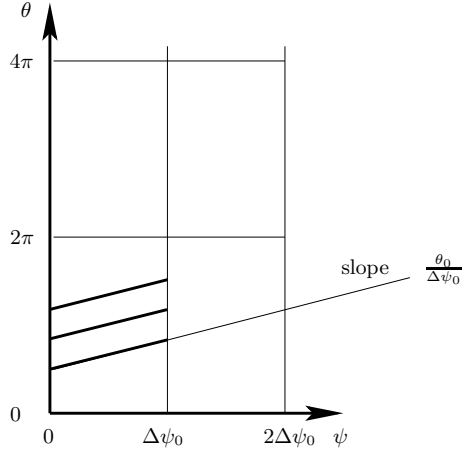}}
\caption{\label{fig: 8} Illustrating the evolution of the angular variables in the representation (\ref{eq: 4.6}), $\theta $ and $\Delta \psi$, as the contour label $\psi_0$ is varied. The periods of these variables are $2\pi$ and $\Delta \psi_0$ respectively, and (\ref{eq: 4.4}) implies that the slope of the line is $\theta_0/\Delta \psi_0$. The evolution can be \lq folded' into a unit cell, and provided $\theta_0/2\pi$ is an irrational number this reduced dynamics is ergodic.
}
\end{figure}

In the limit as $t\to \infty$ the periods associated with varying $\psi$, namely $\Delta \psi_1$ and $\Delta \psi_0$ respectively, both approach zero. As the contour $\psi_0$ is varied, the values of $\theta$ and $\Delta \psi$ both change linearly, along a trajectory illustrated in figure \ref{fig: 8}. The local averaging of the orientation distribution is effected by averaging along this trajectory. Because of the periodicity, the trajectory can be \lq folded back' into a single unit cell. The folded trajectory will fill this unit cell provided $\theta_0/2\pi$ is an irrational number (that is, not a ratio of two integers). Because rational numbers are a measure zero case, we may perform the local average by averaging (\ref{eq: 4.6}) over the unit cell in figure \ref{fig: 8}.

\begin{figure}
\centerline{\includegraphics[width=9cm]{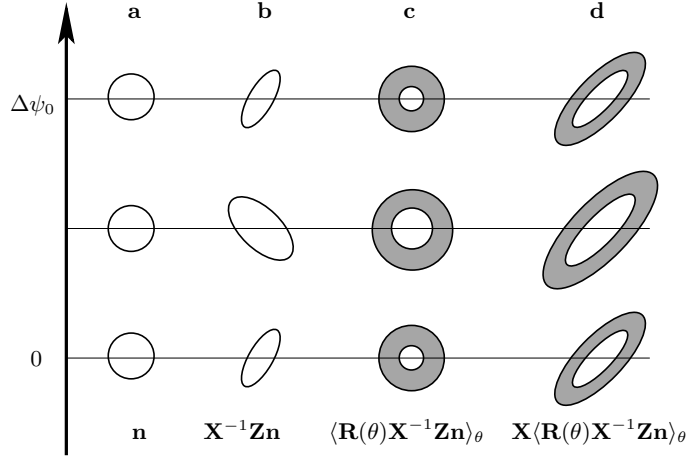}}
\caption{\label{fig: 9} Illustrating the transformations which are applied in succession to the distribution of the initial direction vector ${\bf n}$, in order to produce the vector $\mbox{\boldmath$d$}={\bf M}\,{\bf n}$, where ${\bf M}$ is expressed in the form (\ref{eq: 4.6}). The vector ${\bf n}$ is initially randomly distributed around a unit circle ({\bf a}). After application of the transformation ${\bf X}^{-1}\,{\bf Z}$, this circle is transformed into an ellipse, with the parameters of the ellipse depending periodically upon $\Delta \psi$ ({\bf b}). If we average over the rotation angle of the matrix ${\bf R}(\theta)$, the vectors which are randomly distributed on an elliptical curve are mapped into an annular region ({\bf c}). This region is mapped into an elliptic annulus by the final transformation ${\bf X}$ ({\bf d}).
}
\end{figure}

Consider the behaviour of the order parameter of the locally averaged orientation in terms of the representation (\ref{eq: 4.6}) (without loss of generality  we may consider the line $\tau=0$). We consider a region which is large compared to both of the periods $\Delta \psi_0$ and $\Delta \psi_1$ (note that both periods approach zero in the long-time limit, so this region can be made arbitrarily small). The orientation of ${\bf n}$ is initially distributed randomly around the unit circle. The matrix ${\bf X}^{-1}\,{\bf Z}(\Delta \psi/\Delta \psi_0)$ maps this circle to an ellipse, the parameters of which depend periodically upon $\Delta\psi$, with period $\Delta \psi_0$ (this is illustrated schematically in figure \ref{fig: 9}{\bf a},{\bf b}). We will average over the period $\Delta \psi_0$ as the final stage of our argument. This ellipse is rotated by the angle $\theta$, which depends increasingly sensitively on $\psi$ in the long-time limit, with a period $\Delta \psi_1$ which is inversely proportional to time, so that we can average over the rotation angle $\theta$. Upon averaging over $\theta$, the ellipse is therefore transformed into a circularly symmetric distribution in the plane, as illustrated in figure \ref{fig: 9}{\bf c}. The action of the matrix ${\bf X}$ transforms this annular region into a region bounded by two similar ellipses; see figure \ref{fig: 9}{\bf d}. These have an aspect ratio $\nu$ which is the square root of the ratio of the eigenvalues of ${\bf X}{\bf X}^{\rm T}$, as described in section \ref{sec: 3}. The arguments developed in section \ref{sec: 3} show that the angular distribution $P(\theta)$ depends only upon the aspect ratio of the ellipse, and not upon its overall scale. Furthermore, although the radial distribution in the circular region depends upon $\Delta \psi$, it is only the aspect ratio of the elliptic region which matters, and this is determined solely by the matrix ${\bf X}{\bf X}^{\rm T}$, so that the average over $\Delta \psi$ is trivial.

\begin{figure}
\centerline{\includegraphics[width=12cm]{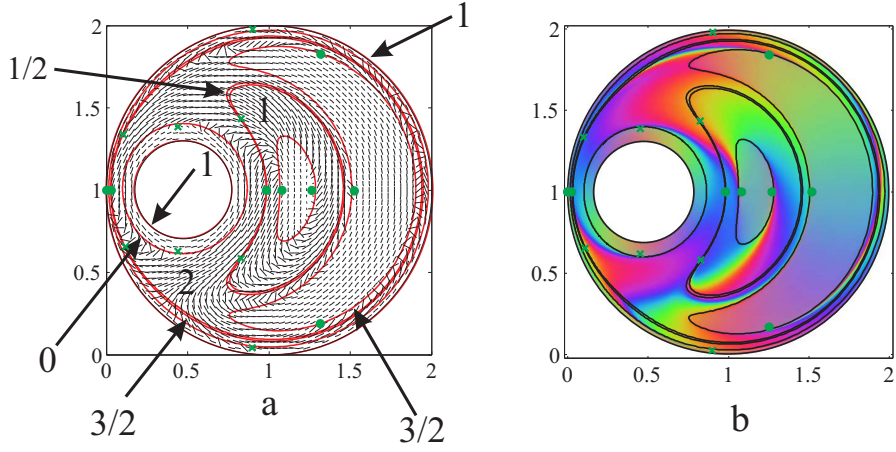}}
\caption{\label{fig: 10}{\bf a} Illustrating the locally averaged order parameter field computed using (\ref{eq: 4.3}) and (\ref{eq: 5.1}). The hyperbolic and elliptic bands are separated by red lines, and the positions of zeros of the order parameter field are indicated by green dots for zeros with Poincar\'e index equal to $\frac{1}{2}$, green crosses for zeros with index $-\frac{1}{2}$.  {\bf b} Is the same image as figure \ref{fig: 4}, with the boundaries between hyperbolic and elliptic bands indicated by solid black lines (and the positions of zeros of $\langle\mbox{\boldmath$\zeta$}\rangle$ are also marked).
}
\end{figure}

Thus we conclude that in the elliptic regions the locally averaged order parameter approaches a limit which varies smoothly as a function of position. The averaged order parameter $\langle \mbox{\boldmath$\zeta$}\rangle(\mbox{\boldmath$r$})$ is determined by the matrix ${\bf X}(\mbox{\boldmath$r$})$ which occurs in the definition of the normal form (\ref{eq: 4.3}), in the same manner as the un-averaged order parameter $\mbox{\boldmath$\zeta$}(\mbox{\boldmath$r$},t)$ is determined from the pseudomonodromy matrix ${\bf M}(\mbox{\boldmath$r$},t,t_0)$. In particular, the equation (\ref{eq: 3.9}) for the  matrix ${\bf K}$ defining the quadratic form for the inertia tensor of the angle distribution is replaced by
\begin{equation}
\label{eq: 5.1}
{\bf K}^{-1}={\bf X}\,{\bf X}^{\rm T}
\ .
\end{equation}
The locally-averaged order parameter $\langle\mbox{\boldmath$\zeta$}\rangle$ points in the direction of the eigenvector corresponding to the largest eigenvalue of ${\bf X}\,{\bf X}^{\rm T}$, and if the square root of the ratio of eigenvalues of this matrix is $\mu$, then $\vert\langle\mbox{\boldmath$\zeta$}\rangle\vert=(\mu-1)/(\mu+1)$. Equation (\ref{eq: 5.1}) is one of the principal results of this paper, since it expresses the long-time limit of the alignments of particles in terms of the normal-form of the transfer matrix ${\bf M}_0$. The locally averaged order parameter field is illustrated in figure \ref{fig: 10}{\bf a} for the same journal bearing example as figures \ref{fig: 4} and \ref{fig: 6}.

\subsection{Continuity of the averaged order parameter}
\label{sec: 5.2}

In a hyperbolic band, where the particles approach a fixed alignment, the asymptotic rod direction at any point $\mbox{\boldmath$r$}$ approaches the dominant eigenvector $\mbox{\boldmath$u$}_+$ of the transfer matrix ${\bf M}_0(\mbox{\boldmath$r$})$, for a periodic orbit which ends at $\mbox{\boldmath$r$}$. In the long-time limit, a local average of this order parameter field, $\langle\mbox{\boldmath$\zeta$}\rangle(\mbox{\boldmath$r$})$, is also a smooth function of position throughout the elliptic region. We should consider whether the locally averaged order parameter varies continuously upon passing between elliptic and hyperbolic regions.

In \cite{Wil+08}, we showed that the transfer matrix at a boundary between elliptic and hyperbolic regions where ${\rm tr}({\bf M})=2$ is in the form of a generalised shear:
\begin{equation}
\label{eq: 5.2}
{\bf M}={\bf R}(\phi)\,{\bf S}(\kappa)\,{\bf R}(-\phi)
\end{equation}
where ${\bf R}(\phi)$ is a rotation matrix and ${\bf S}(\kappa)$ is a shear of the form
\begin{equation}
\label{eq: 5.3}
{\bf S}(\kappa)=\left(\begin{array}{cc}1&\kappa \cr 0 & 1\cr \end{array}\right)\ .
\end{equation}
It follows that eigenvectors of the monodromy matrix become co-linear as we approach the boundary between elliptic and hyperbolic regions: both eigenvectors approach $\mbox{\boldmath$u$}=(\cos\phi,\sin\phi)$, while both eigenvalues approach unity. As we approach such a boundary from the hyperbolic side, the order parameter field aligns with this common eigenvector. It will prove useful to express (\ref{eq: 5.2}) in component form: introducing the notations $c=\cos\phi$, $s=\sin\phi$, we find that
\begin{equation}
\label{eq: 5.4}
{\bf M}=\left(\begin{array}{cc}
1+\kappa cs & \kappa c^2 \cr
-\kappa s^2 & 1-\kappa cs
\end{array}\right)
\ .
\end{equation}

However, it is not immediately clear what happens as we approach the boundary from the elliptic side, where the transfer matrix can be expressed in the form (\ref{eq: 4.3}). As the boundary is approached, the angle $\theta_0$ in (\ref{eq: 4.3}) approaches zero, because ${\rm tr}({\bf M})=2\cos \theta_0\to 2$, and in the following discussion we treat $\theta_0$ as a small number. It is clear that the matrix ${\bf X}$ in the representation (\ref{eq: 4.3}) must become singular in order to approach (\ref{eq: 5.2}) as $\theta_0\to 0$. Let us assume that in this limit ${\bf X}$ takes the form:
\begin{equation}
\label{eq: 5.5}
{\bf X}=
\left(\begin{array}{cc}
\cos(\phi+\delta \phi) & \cos(\phi-\delta \phi) \cr
\sin(\phi+\delta \phi) & \sin(\phi-\delta \phi)
\end{array}\right)
=
\left(\begin{array}{cc}
c-\delta\phi\ s & c+\delta\phi s \cr
s+\delta\phi c & s-\delta\phi c
\end{array}\right)
+O(\delta \phi^2)
\end{equation}
where we use the notations $c=\cos(\phi)$, $s=\sin(\phi)$, and where we shall assume that the small change in the angle is
\begin{equation}
\label{eq: 5.6}
\delta \phi=\frac{\theta_0}{\kappa}+O(\theta_0^2)
\ .
\end{equation}
We find ${\rm det}({\bf X})=2\theta_0$, so the assumed form for ${\bf X}$ does indeed become singular as $\theta_0\to 0$. Inserting the {\em ansatz} (\ref{eq: 5.5}), (\ref{eq: 5.6}) into (\ref{eq: 4.3}), approximating
\begin{equation}
\label{eq: 5.7}
{\bf R}(\theta_0)={\bf I}+\theta_0{\bf J}+O(\theta_0^2)\ ,\ \ \
{\bf J}=\left(\begin{array}{cc}
0 & 1 \cr
-1 & 0
\end{array}\right)
\end{equation}
and ignoring $O(\theta_0^2)$ terms, we find:
\begin{equation}
\label{eq: 5.8}
{\bf M}=\left(\begin{array}{cc}
1+\kappa cs  & -\kappa c^2  \cr
\kappa s^2  & 1-\kappa cs
\end{array}\right)
+\theta_0 \left(\begin{array}{cc}
c^2-s^2  & 0 \cr
0  & s^2-c^2
\end{array}\right)
+O(\theta_0^2)
\ .
\end{equation}
In the limit as $\theta_0\to 0$ we find that this expression agrees with (\ref{eq: 5.4}), which confirms that the ansatz (\ref{eq: 5.5}), (\ref{eq: 5.6}) was correct. We can now use this expression for ${\bf X}$ in equation (\ref{eq: 5.1}) to calculate the form of the matrix ${\bf K}^{-1}$ defining the quadratic form characterising the order parameter: we obtain
\begin{equation}
\label{eq: 5.9}
{\bf K}^{-1}=2\left(\begin{array}{cc}
c^2  & cs   \cr
cs   & s^2
\end{array}\right)
+O(\theta_0^2)
\ .
\end{equation}
The term which is independent of $\theta_0$ is a singular matrix: its eigenvectors are $\mbox{\boldmath$U$}_1=(\cos\phi,\sin\phi)$ with eigenvalue $\Lambda_1=1$, and $\mbox{\boldmath$U$}_2=(\sin\phi,-\cos\phi)$ with eigenvalue $\Lambda_2=0$. This shows that in the limit as $\theta_0\to 0$ the ellipse which is defined by the quadratic form $({\bf X}\,{\bf X}^{\rm T})^{-1}$ degenerates into a line, which is aligned with the common eigenvector of (\ref{eq: 5.2}). Because the aspect ratio of the ellipse approaches infinity, the modulus of the order parameter approaches unity as the boundary is approached. We conclude that the locally averaged order parameter is continuous at the boundary between elliptic and hyperbolic regions (although it clearly has discontinuous derivatives).

Finally we comment on the nature of the discontinuity of the order parameter at the boundary between the elliptic and hyperbolic bands. The fact that the $O(\theta_0)$ term in (\ref{eq: 5.9}) is equal to zero implies that the determinant of ${\bf X}\,{\bf X}^{\rm T}$ is ${\rm det}({\bf K}^{-1})=O(\theta_0^2)$, implying that $\Lambda_2=O(\theta_0^2)$. This implies that the aspect ratio of the ellipse is $\nu \sim \theta_0^{-2}$. Because ${\rm tr}({\bf M})=2\cos\theta_0$ has a linear dependence upon the distance $d$ from the boundary with the hyperbolic region, we conclude that $\theta_0\sim \sqrt{d}$, so that $\nu \sim 1/d$. This in turn implies that the magnitude of the order parameter approaches unity  linearly upon approaching the boundary of an elliptic band, implying that $\langle\mbox{\boldmath$\zeta$}\rangle(\mbox{\boldmath$r$})$ has a discontinuous first derivative.

\subsection{Poincar\'e indices of the order parameter}
\label{sec: 5.3}

The Poincar\'e index must be the same for any curve lying in a hyperbolic band, because $\mbox{\boldmath$u$}_+$ cannot have any singularities there \cite[]{Wil+08}. The Poincar\'e index is most efficiently determined by evaluating $\mbox{\boldmath$u$}_+$ around a given contour of $\psi$ within the hyperbolic band. These Poincar\'e indices are indicated for each of the hyperbolic bands in figure \ref{fig: 6}\,{\bf b}. They were also evaluated by \cite{Sze93}: his \lq flip numbers' are twice the Poincar\'e index.

We have seen that each hyperbolic band is associated with a Poincar\'e index, and simulations confirm that the Poincar\'e indices of different hyperbolic bands need not be equal. We have also seen that $\langle \mbox{\boldmath$\zeta$}\rangle$ is continuous everywhere, so that a Poincar\'e index can also be ascribed to the averaged order parameter field $\langle\mbox{\boldmath$\zeta$}\rangle(\mbox{\boldmath$r$})$ in the elliptic bands. This raises the following question: is there a rule for determining the difference between the Poincar\'e indices of the hyperbolic bands in terms of a property of the intervening elliptic band?

The analogy with Bloch bands in solid-state physics suggests that a rule for Poincar\'e indices might be found. The wavefunction of a Bloch band at any given energy is characterised by a Bloch wavevector $k$, such that on traversing one period $L$ of the potential the wavefunction accumulates a phase factor $\exp({\rm i}kL)$. The phase $\theta_0$ in (\ref{eq: 4.3}) corresponds to $kL$ in the Bloch wavefunction. The wavevector is related to the monodromy matrix by $\vert{\rm tr}{\bf M}\vert=2\cos(kL)$. On traversing a band, the wavefunction therefore rotates by $\pi$ for every period of the potential. By analogy, in a steady flow we might expect that the axis rotates by $\pm \pi$ on crossing every elliptic band, which would imply that the Poincar\'e index changes by $\pm\frac{1}{2}$ on crossing every elliptic band. The following argument shows that this physical intuition is partially correct.

\begin{figure}
\centerline{\includegraphics[width=6cm]{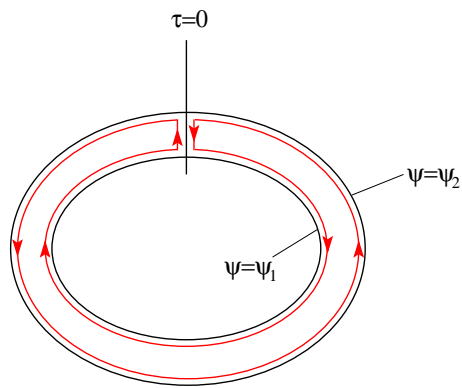}}
\caption{\label{fig: 11} Illustrating the path used in the discussion of Poincar\'e indices in section \ref{sec: 5.3}.
}
\end{figure}

A rule for changes of the Poincar\'e index is obtained by the following argument. We assume that the elliptic region has a simple annular topology, although cases where an elliptic region has two or more \lq holes' occur. In the elliptic band the eigenvalues of ${\bf M}_0$ are complex numbers, with both the eigenvalues and eigenvectors occurring as complex conjugate pairs. We can multiply the eigenvectors by complex numbers chosen so that these vectors are purely real at the band edges. Let us combine the two eigenvectors $\mbox{\boldmath$u$}_1$, $\mbox{\boldmath$u$}_2=\mbox{\boldmath$u$}_1^\ast$ to yield a real-valued vector $\mbox{\boldmath$a$}=\frac{1}{2}[\mbox{\boldmath$u$}_1+\mbox{\boldmath$u$}_2]$. This vector depends upon position, because the matrix ${\bf M}_0$ depends upon the position $\mbox{\boldmath$r$}$. In in the following we use the same coordinates as in section \ref{sec: 4}, so that positions within the band are labelled by the value of the stream function contour, $\psi_0$, and by the time $\tau$ taken to reach the point from a specified starting point on the contour (see figure \ref{fig: 7}). Note that the matrix ${\bf M}_0$ at different points around the contour $\psi_0$ is related by a similarity transformation: ${\bf M}_0(\psi_0,\tau)={\bf M}(\psi_0,\tau)\,{\bf M}_0(\psi_0,0)\,{\bf M}^{-1}(\psi_0,\tau)$, implying that eigenvectors satisfy $\mbox{\boldmath$u$}_i(\psi_0,\tau)={\bf M}(\psi_0,\tau)\mbox{\boldmath$u$}_i(\psi_0,0)$. Now let us consider some properties of the vector field
\begin{equation}
\label{eq: 5.10}
\mbox{\boldmath$A$}(\psi_0,\tau)={\bf M}(\psi_0,\tau)\mbox{\boldmath$a$}(\psi_0,0)=\frac{1}{2}{\bf M}(\psi_0,\tau)[\mbox{\boldmath$u$}_1(\psi_0,0)+\mbox{\boldmath$u$}_2(\psi_0,0)]
\ .
\end{equation}
We note the following properties of this vector field:

\begin{enumerate}

\item At the inner and outer edges of the elliptic band (we label these contours $\psi_1$ and $\psi_2$ respectively), the two eigenvectors $\mbox{\boldmath$u$}_1$, $\mbox{\boldmath$u$}_2$ become colinear, and the real-valued vector $\mbox{\boldmath$A$}(\psi,\tau)$ corresponds to the single eigenvector of the monodromy matrix. The vector $\mbox{\boldmath$A$}$ therefore corresponds to the long-time limit of the order parameter at the inner and outer edges of the elliptic band, and the Poincar\'e index of $\mbox{\boldmath$A$}$ on the inner and outer edges corresponds to the Poincar\'e indices ($N_1$ and $N_2$ respectively) of the surrounding hyperbolic bands.

\item The vector field $\mbox{\boldmath$A$}(\psi_0,\tau)$ is clearly a smooth function of position within the elliptic band. Also, because ${\bf M}(\psi_0,\tau)$ is non-singular, and the vector $\mbox{\boldmath$a$}(\psi_0,0)$ does not vanish for any value of $\psi_0$ in the interval $[\psi_1,\psi_2]$, this vector field $\mbox{\boldmath$a$}(\psi_0,\tau)$ has no zeros in the elliptic band.

\item Let us consider a closed curve which is composed of the line $\tau=0$ traversed from the outer edge to the inner (from $\psi_0=\psi_2$ to $\psi_0=\psi_1$), the inner edge of the elliptic band (that is, the line $\psi_0=\psi_1$) traversed clockwise around one period, the line $\tau=0$ traversed from $\psi_0=\psi_1$ to the outer edge $\psi_0=\psi_2$, and then the outer edge (the line $\psi_0=\psi_2$) traversed counterclockwise back to the starting point. This path is illustrated in figure \ref{fig: 11}. Because the vector field $\mbox{\boldmath$A$}$ has no zeros and is everywhere smooth within this region, the Poincar\'e index $N$ of this field evaluated on the specified path is equal to zero.

\item However, we note that the vector field $\mbox{\boldmath$A$}(\psi,\tau)$ is periodic on the segments which correspond to the inner and outer edges of the elliptic band ($\psi_0=\psi_1$ or $\psi_0=\psi_2$), so that we can talk about a Poincar\'e index defined on these segments of the path in isolation. Furthermore, because $\mbox{\boldmath$A$}$ corresponds to the order parameter field $\langle \mbox{\boldmath$\zeta$} \rangle$ at the band edges, we see that the contribution to the Poincar\'e index $N$ of $\mbox{\boldmath$A$}$ on the closed paths which arise from the inner and outer band edges is equal to the difference between the Poincar\'e index of the order parameter field at the inner and outer edges of the elliptic band. The can therefore deduce this difference (that is, $N_1-N_2$) by evaluating the contribution to the Poincar\'e index which arises from the two segments along the line $\tau=0$.

\item Although the path in space which is followed by the two \lq radial' segments of path considered in 3 above is the same, the vector field differs because in one case the matrix ${\bf M}(\psi_0,T)={\bf M}_0(\psi_0,0)$ has been applied to the vector $\mbox{\boldmath$a$}(\psi_0,0)$. This vector is constructed from the two complex-conjugate eigenvectors of ${\bf M}_0$, for which the corresponding eigenvalues may be written as $\exp({\rm i}K)$, where ${\rm tr}({\bf M}_0)=2\cos(K)$. If we write the eigenvectors of ${\bf M}_0$ in the form $\mbox{\boldmath$u$}=\mbox{\boldmath$a$}+{\rm i}\mbox{\boldmath$b$}$, where $\mbox{\boldmath$b$}$ is a real-valued vector, then we can express the relation between the vector $\mbox{\boldmath$A$}$ on the two radial components of the closed path as follows:
\begin{equation}
\label{eq: 5.11}
\mbox{\boldmath$A$}(\psi_0,T(\psi_0))=\cos(K(\psi_0))\mbox{\boldmath$a$}(\psi_0,0)+\sin(K(\psi_0))\mbox{\boldmath$b$}(\psi_0,0)\ .
\end{equation}

\item Equation (\ref{eq: 5.11}) leads to two possible conclusions. The band edges correspond to points at which ${\rm tr}({\bf M}_0)=\pm 2=2\cos(K)$. This implies that $\sin(K)=0$ at the band edges and $\cos(K)=\pm 1$. If ${\rm tr}({\bf M}_0)$ has opposite signs at the two band edges, then equation (\ref{eq: 5.11}) implies that $\mbox{\boldmath$A$}$ changes sign when the two radial elements of the closed path in figure \ref{fig: 11} are traversed in opposite directions. Because the Poincar\'e index for the composite path is equal to zero, this change of sign implies that the Poincar\'e indices of the inner and outer band edges differ by $\pm \frac{1}{2}$. Conversely, if the sign of ${\rm tr}({\bf M}_0)$ is the same on the inner and outer band edges, then the Poincar\'e indices of the inner and outer bands edges are equal. A more detailed argument, requiring information about eigenvectors of ${\bf M}_0$ as well as its trace, is required to establish the sign of the change in the Poincar\'e index.

\end{enumerate}

In the solid-state physics context, the structure of the Schr\"ondinger equation implies that the trace of the monodromy matrix always does change sign upon crossing a band. In the problem we consider here, ${\rm tr}({\bf M}_0)$ need not change sign upon crossing an elliptic band, so that the surmise about the Poincar\'e index based on the solid-state physics analogy is only partially correct.

\subsection{Singularities of the order parameter}
\label{sec: 5.4}

As pointed out in sections \ref{sec: 5.2} and \ref{sec: 5.3} above, the locally averaged order parameter vector $\langle\mbox{\boldmath$\zeta$}\rangle$ varies smoothly and is defined everywhere within the elliptic bands. However we have seen that the Poincar\'e index of the order parameter field may differ by $\pm\frac{1}{2}$ between the inner and outer edges of the band. When these Poincar\'e indices are different, there must be at least one singular point inside the band, where there is a zero of the averaged order parameter vector field $\langle\mbox{\boldmath$\zeta$}\rangle(\mbox{\boldmath$r$})$. We now consider the structure of these singularities. A similar argument is presented in \cite{Bez+09}, where we discuss singularities of the order parameter $\mbox{\boldmath$\zeta$}$ for random flows. Here we discuss singularities of $\langle \mbox{\boldmath$\zeta$}\rangle$ for steady flows, and find that the mathematical structure of the singularities is the same, although the argument has a different structure.

\begin{figure}
\centerline{\includegraphics[width=8cm]{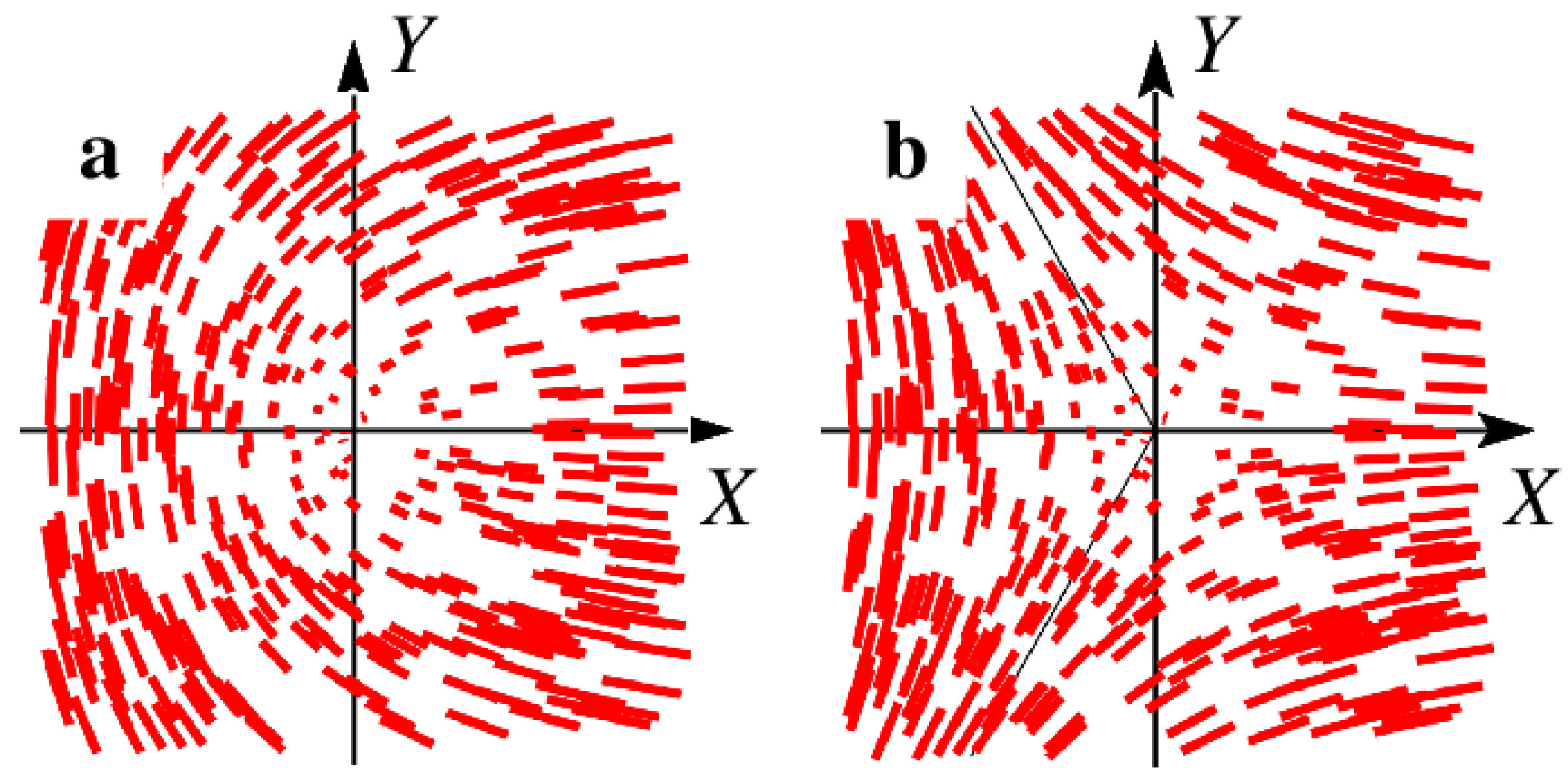}}
\caption{\label{fig: 12} Illustrating the normal forms for the zeros of the locally averaged order parameter field $\langle\mbox{\boldmath$\zeta$}\rangle(X,Y)$: {\bf a} $s=+1$ leads to a {\em core} singularity, {\bf b} $=-1$ leads to a {\em delta} singularity.
}
\end{figure}

In the case where ${\bf X}$ in equation (\ref{eq: 4.6}) is a unit matrix, there is a singularity where the orientation remains uniformly distributed, implying that the order parameter vector vanishes. We now examine the structure of the position-dependence of this matrix in the vicinity of this singular point. A general $2\times 2$ matrix ${\bf A}$ can be written in the form
\begin{equation}
\label{eq: 5.12}
{\bf A}=\alpha\, {\bf R}(\phi)\,{\rm diag}(\lambda,\lambda^{-1})\,{\bf S}(\kappa)
\end{equation}
described by four parameters $\alpha$, $\phi$, $\lambda$, $\kappa$, where ${\bf S}(\kappa)$ is the shear matrix, (\ref{eq: 5.3}). Consider the use of the representation (\ref{eq: 5.12}) to parametrise the matrix ${\bf X}$ in (\ref{eq: 4.3}). First note that because the scaling constant $\alpha$ and the rotation matrix ${\bf R}(\phi)$ both commute with ${\bf R}(\theta)$, if we express ${\bf X}$ in the form (\ref{eq: 5.12}), the values of $\alpha$ and $\phi$ are irrelevant, so that we may write ${\bf X}$ as a member of a two-parameter family: ${\bf X}={\rm diag}(\lambda,\lambda^{-1}){\bf S}(\tau)$. By a linear transformation ${\bf T}$ of the coordinate system, we may represent the position $\mbox{\boldmath$r$}$ in the vicinity of a zero at $\mbox{\boldmath$r$}_0$ in terms of coordinates $\mbox{\boldmath$X$}=(X,Y)$, writing $\mbox{\boldmath$X$}={\bf T}(\mbox{\boldmath$r$}-\mbox{\boldmath$r$}_0)$. This change of coordinates is non-inverting (that is, ${\rm det}({\bf T})>0$) and is determined so that $\lambda=1+\frac{1}{2}X+O(\mbox{\boldmath$X$}^2)$, $\kappa=s Y+O(\mbox{\boldmath$X$}^2)$, with the sign $s=\pm 1$ chosen so that ${\rm det}({\bf T})>0$. The position dependence of the matrix ${\bf X}$ may therefore be parametrised as
\begin{eqnarray}
\label{eq: 5.13}
{\bf X}&=&\left(\begin{array}{cc}
1+\frac{1}{2}X & 0 \cr 0 & 1-\frac{1}{2}X \cr
\end{array}\right)
\left(\begin{array}{cc}
1 & sY \cr 0 & 1\cr
\end{array}
\right)+O(\mbox{\boldmath$X$}^2)
\nonumber \\
&=&\left(\begin{array}{cc}
1+\frac{1}{2}X & sY \cr  0 & 1-\frac{1}{2}X \cr
\end{array}\right)+O(\mbox{\boldmath$X$}^2)\ .
\end{eqnarray}
The parameter dependence of the matrix ${\bf K}^{-1}={\bf X}\,{\bf X}^{\rm T}$ is therefore of the form
\begin{equation}
\label{eq: 5.14}
{\bf K}^{-1}=\left(\begin{array}{cc}
1+X & sY \cr
sY  & 1-X \cr
\end{array}\right)+O(\mbox{\boldmath$X$}^2)
\ .
\end{equation}
This matrix has eigenvalues $\lambda_\pm =1\pm R$, where $R=\sqrt{X^2+Y^2}$, and if we write $(X,Y)=(R\cos \Theta,R\sin \Theta)$, we find that the eigenvector corresponding to the largest eigenvalue, $1+R$, has angle $\theta=s\frac{1}{2}\Theta$. The aspect ratio is $\nu =\sqrt{(1+R)/(1-R)}=1+R+O(R^2)$. The magnitude of the order parameter is then $\vert\mbox{\boldmath$\zeta$}\vert=R/2+O(R^2)$, so that the locally averaged order parameter is
\begin{equation}
\label{eq: 5.15}
\langle \mbox{\boldmath$\zeta$}\rangle(X,Y)=\frac{R}{2}\,{\bf n}(\frac{s}{2}\Theta)+O(\mbox{\boldmath$X$}^2)
\ .
\end{equation}
The field $\langle\mbox{\boldmath$\zeta$}\rangle(X,Y)$ is illustrated in figure \ref{fig: 12} for both choices of the sign $s$. In both cases the normal form of the singularity, (\ref{eq: 5.15}), resembles forms which are seen in ridge patterns of fingerprints (first described by \cite{Hen00}): we have a {\em core} singularity when $s=+1$ or a {\em delta} singularity when $s=-1$.

The singularities of our order parameter field are very closely related to \lq umbilic points' on surfaces, where the height $z$ above the Cartesian plane is $z=f(x_1,x_2)$. An umbilic point is a point where the magnitudes of the principal curvatures are equal, so that the surface is locally isotropic. Different ways of categorising umbilic points are discussed by \cite{Ber+77}. The principal curvatures are the defined by the eigenvalues and eigenvectors of the real symmetric Hessian matrix, with elements $\partial^2 f/\partial x_i\partial x_j$. This is analogous to considering the matrix ${\bf K}(x,y)$ discussed above, and the direction of one of the principal axes of curvature has the same singularities as the direction of our order parameter. The classification of directions of principal curvatures discussed by \cite{Ber+77} lists three types of singularity, {\em star}, {\em lemon} and {\em monstar}. The star is equivalent to the {\em delta} singularity of fingerprint patterns. The lemon and monstar are subdivisions of the {\em core} singularity. They are distinguished by the number of lines along which the vector field is aligned radially. In the core and delta singularities illustrated in figure \ref{fig: 12}, there is one such line for the core singularity and there are three such lines for the delta singularity. This figure illustrates the two singularities expressed in the normal form coordinates, $(X,Y)$. Upon transforming back to the original Cartesian coordinates, $\mbox{\boldmath$x$}={\bf T}^{-1}\,\mbox{\boldmath$X$}$, however, angles need to be preserved, and for some choices of ${\bf T}$ the core singularity has two additional lines where the vector field points radially. Core singularities with one radial line are termed {\em lemons} in \cite{Ber+77}, and those with three radial lines are {\em monstars}. \cite{Den08} gives a clear and nicely illustrated discussion of monstar singularities.

Zeros of the order parameter can be identified in the journal bearing example and their positions are plotted in figure \ref{fig: 10}. These are generic zeros with the same structure as the normal forms discussed above, but the transformation ${\bf T}$ from the original coordinate system to that of the normal forms is close to being singular, so that the structure of the normal forms is highly distorted in figure \ref{fig: 10}. The particles in contact with the moving walls tumble, but their order parameter is in alignment with the walls at the boundary. The Poincar\'e index of both the inner and outer boundaries is therefore $+1$, like that of the \lq vortex' in figure \ref{fig: 2}\,{\bf a}. The Poincar\'e index of the hyperbolic regions can be seen to obey the rule discussed in section \ref{sec: 5.3}, changing by no more than $\pm \frac{1}{2}$ on crossing each annular elliptic band. There are a total of sixteen zeros of the order parameter lying in the elliptic bands: eight cores and eight deltas. Their topological charges can be seen to be consistent with the changes of the Poincar\'e index on crossing elliptic bands. It is interesting to note that in figure \ref{fig: 10} the set of zeros  is symmetric under reflection, despite the fact that the order parameter field is not. We discuss the behaviour under reflection in section \ref{sec: 5.6} below.

\begin{figure}
\centerline{\includegraphics[width=13cm]{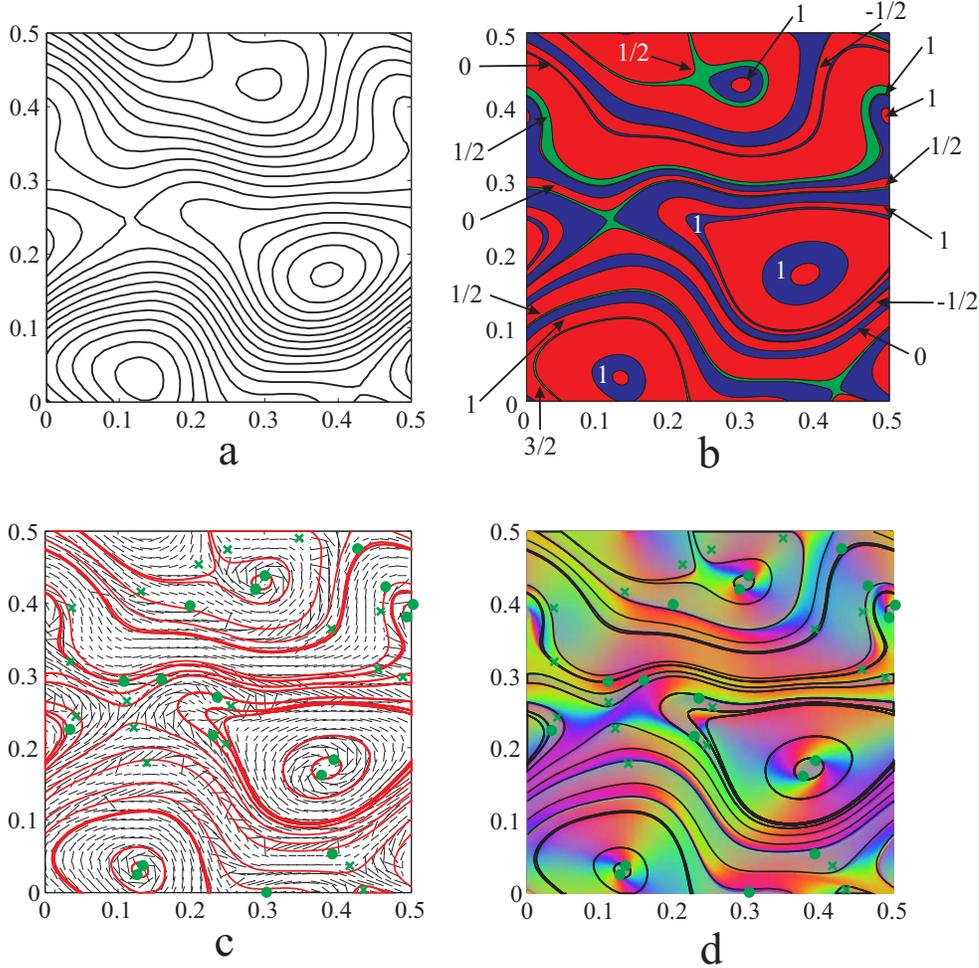}}
\caption{\label{fig: 13} Illustrating particle alignment in a randomly-generated stream function: {\bf a} Contours of the
stream function, which is periodic on a square of length $\frac{1}{2}$. {\bf b} Shows the elliptic bands, $\vert {\rm tr}({\bf M}_0)\vert<2$ (red), and hyperbolic bands, ${\rm tr}({\bf M}_0)<-2$ (green) and ${\rm tr}({\bf M}_0)>2$ (blue).
The hyperbolic bands are labelled with their Poincar\'e index. {\bf c} Shows the locally-averaged order parameter field in the long-time limit. {\bf d} Shows the colour mapping of the locally-averaged long-time order parameter field. The zeros are marked with green dots (cores) and crosses (deltas). In this figure the aspect ratio parameters of the rod-like particles are $\alpha_1=0.875$, $\alpha_2=0.125$, $\beta=\sqrt{7}$.
}
\end{figure}

We also investigated the alignment of particles for a \lq generic' stream function, defined by a two-dimensional real-valued Fourier series on a square domain with random Fourier coefficients. An example is shown in figure \ref{fig: 13}, which shows contours of the stream function ({\bf a}), regions where the transfer matrix is hyperbolic and elliptic and Poincar\'e indices of the hyperbolic bands ({\bf b}), the averaged order parameter field ({\bf c}) and its colour mapping ({\bf d}), with the zeros marked. It can be verified that this example satisfies the rule discussed in section \ref{sec: 5.3}, which related the Poincar\'e indices to ${\rm tr}({\bf M}_0)$.

\subsection{Behaviour close to centres of rotation}
\label{sec: 5.5}

The arguments in section \ref{sec: 5.4} above show how the existence of zeros of the order parameter may be deduced in elliptic bands which lie between hyperbolic bands. We now discuss the regions surrounding stable fixed points of the fluid flow.

The fluid flow has elliptic fixed points (centres of rotation) at maxima and minima of the stream function $\psi(x,y)$.  Szeri analysed the motion of rods at these elliptic fixed points \cite[]{Sze93}. He showed that the rod axis rotates at the fixed point, with a frequency which is less than the frequency at which fluid elements rotate at this point. This can be seen immediately from the general solution (\ref{eq: 2.6}), because at the fixed point the pseudomonodromy matrix is constant in time, so that the solution discussed in section \ref{sec: 2} can be applied directly. It follows that the elliptic fixed points of the flow are always surrounded by elliptic bands of the pseudomonodromy matrix.

It is natural to ask whether the stable fixed points of the fluid flow correspond to zeros of the order parameter. At the stable fixed point, the pseudomonodromy matrix ${\bf M}$ is generated by exponentiating a constant velocity gradient ${\bf A}$, so that the transfer matrix is ${\bf M}_0=\exp({\bf B}T)$, where $T$ is limit of the period of the flow fluid as the fixed point is approached, and ${\bf B}$ is the matrix defined by (\ref{eq: 2.2}), evaluated at the fixed point. The normal-form decomposition of ${\bf M}_0$ will be a pure rotation if the minimum or maximum is (to leading order) circularly symmetric, but in the general the matrix ${\bf X}$ which occurs in (\ref{eq: 4.3}) will not be the identity matrix. Thus we see that, except where fixed points are isotropic, the order parameter is non-zero at stable fixed points of the velocity field.

Inspection of figures \ref{fig: 10} and \ref{fig: 13} confirms that the stable fixed points always occur in regions where the transfer matrix is elliptic, and that stable fixed points do not coincide with zeros of the order parameter.

\subsection{Reflection symmetry}
\label{sec: 5.6}

The journal bearing example which is illustrated in figures \ref{fig: 4}, \ref{fig: 6} and \ref{fig: 10} above has a stream function which is invariant under reflection about the line $y=0$. It is interesting to consider the extent to which this symmetry is reflected in the alignment of the rod-like particles. This is most easily understood by comparing the transfer matrix ${\bf M}_0$ at a point $\mbox{\boldmath$r$}=(x,y)$ with its value ${\bf M}_0^{\rm R}$ at a reflected point $\mbox{\boldmath$r$}^{\rm R}=(x,-y)$. We find it convenient to represent the effect of the reflection by a matrix $\mbox{\boldmath$\Sigma$}$:
\begin{equation}
\label{eq: 5.16}
\mbox{\boldmath$r$}^{\rm R}=\mbox{\boldmath$\Sigma$}\,\mbox{\boldmath$r$}
\ ,\ \ \
\mbox{\boldmath$\Sigma$}=\left(\begin{array}{cc}
1 & 0 \cr
0 & -1
\end{array}\right)
\ .
\end{equation}
The sense of rotation (clockwise or counter-clockwise) about a contour of the stream function is reversed under reflection, which corresponds to taking the inverse of the pseudomonodromy matrix. We can therefore construct ${\bf M}_0^{\rm R}$ by applying a reflection, applying time-reversed propagation at $\mbox{\boldmath$r$}$, and then reflecting again, that is
\begin{equation}
\label{eq: 5.17}
{\bf M}_0^{\rm R}=\mbox{\boldmath$\Sigma$}\,{\bf M}_0\,\mbox{\boldmath$\Sigma$}
\ .
\end{equation}
In component form, the elements two transfer matrices are therefore related as follows:
\begin{equation}
\label{eq: 5.18}
{\bf M}_0=\left(\begin{array}{cc}
m_{11} & m_{12} \cr
m_{21} & m_{22}
\end{array}\right)
\ ,\ \ \
{\bf M}_0^{\rm R}=\left(\begin{array}{cc}
m_{22} & m_{12} \cr
m_{21} & m_{11}
\end{array}\right)
\ .
\end{equation}
The corresponding matrices describing the quadratic form for the time-averaged order parameter, ${\bf K}^{-1}={\bf X}\,{\bf X}^{\rm T}$ at $\mbox{\boldmath$r$}$ and $({\bf K}^{\rm R})^{-1}$ at $\mbox{\boldmath$r$}^{\rm R}$ are therefore related as follows:
\begin{equation}
\label{eq: 5.19}
{\bf K}^{-1}=\left(\begin{array}{cc}
k_{11} & k_{12} \cr
k_{12} & k_{22}
\end{array}\right)
\ ,\ \ \
({\bf K}^{\rm R})^{-1}=\left(\begin{array}{cc}
k_{11} & -k_{12} \cr
-k_{12} & k_{22}
\end{array}\right)
\ .
\end{equation}
Since they have the same determinant and trace, they have the same
eigenvalues and the length of the order parameter vector is the same at the reflected point. The direction of the order parameter at two points satisfies $\theta+\theta^{\rm R}=2\pi$, because the signs of off-diagonal elements are
opposite. This relation between the directions at reflected points is apparent in figure \ref{fig: 10}{\bf a}.

It also follows that zeros of the order parameter come in symmetric pairs, except where there is a zero on the axis of symmetry. Also, the Poincar\'e index of a zero and its reflected partner must be the same. This is confirmed by inspection of figure \ref{fig: 10}{\bf b}. A further consequence is that the directions on the symmetry axis can only be aligned with the axis, or else perpendicular.


\section{Concluding remarks}
\label{sec: 6}

The alignment of small anisotropic particles due to velocity gradients in fluid flows is a significant problem, with a broad range of potential applications. There is as yet no general solution for a triaxial body in a three-dimensional flow, but the special case of an axisymmetric body is expected to exhibit most of the physically important phenomena. A simple and powerful general solution for the axisymmetric case was given by \cite{Sze93}, who showed how the orientation may be obtained from a companion linear problem. This present work is the third of three papers which have investigated the consequences of this solution in different situations. These concluding remarks will set the results of this paper in context with our earlier work.

The characteristics of the solution depend upon whether the flow is chaotic or recirculating, and upon whether we average over a random initial configuration of the particles. In \cite{Wil+08}, we considered the case where the initial orientation of the particles is not random, and where they are advected in a non-steady flow (which may assumed to have a positive Lyapunov exponent in most cases). We showed that the particle orientation field ${\bf n}(\mbox{\boldmath$r$},t)$ is, strictly speaking, a smooth function of the position $\mbox{\boldmath$r$}$, but that numerical simulations in two dimensions exhibit apparent singularities, which resemble the core and delta singularities in the ridge patterns of fingerprints. We showed how the occurrence of these apparent singularities can be explained. We also discussed the behaviour of the solution (\ref{eq: 2.6}) in the long-time limit: we showed that, despite the increasing sensitivity of ${\bf M}(\mbox{\boldmath$r$},\mbox{\boldmath$r$}_0,t)$ to the final position $\mbox{\boldmath$r$}$, the direction vector field is statistically stationary in the long-time limit. This is an apparently paradoxical conclusion, in that under the assumption that a Lyapunov exponent is positive, we showed that there is {\em not} increasing sensitivity to the initial condition in the long-time limit.

For non-steady flows, in the long-time limit the pseudomonodromy matrix becomes hyperbolic almost everywhere, and the particles align with its dominant eigenvector. This implies that in the long-time limit, the initial condition is forgotten for random velocity fields. However there are cases when the pseudomonodromy matrix does not have large eigenvalues, so that there is some memory of the initial orientation. In these cases, it is usually physically appropriate to assume that the initial particle directions are randomly distributed, and to average over a uniform distribution of the initial angle. In these cases the typical orientation of the particles is described by an order parameter field, $\mbox{\boldmath$\zeta$}(\mbox{\boldmath$r$},t)$. The order parameter is required to characterise the direction field at short times, for any type of flow. Also, in the case of recirculating flows, such as those considered here, there is no guarantee that the pseudomonodromy matrix has an eigenvalue which increases with time. In \cite{Bez+09} we considered the order parameter field for random flows at short times, and showed that this field has {\em true} singularities, which resemble the core and delta singularities of fingerprints. Their normal forms were characterised, and their existence was demonstrated experimentally.

In this paper we considered the second situation where the order parameter is relevant, that of a recirculating flow in the long-time limit. This example turns out to be more subtle than the case of random flows at short times. This is primarily because it is a complement to the theorem proved in \cite{Wil+08}. In this case the Lyapunov exponent of the flow is zero, and the pattern formed by the order parameter field {\em can} depend increasingly sensitively on position as $t\to \infty$, (witnessed by the increasingly tight spirals shown in figure \ref{fig: 1}). In order to fully understand the evolution of the order parameter, in such cases it is necessary to understand how to compute the order parameter of the locally averaged orientation in the long-time limit. It is this calculation which is the central achievement of the present work. The result is contained in equations (\ref{eq: 4.3}) and (\ref{eq: 5.1}), which show how the matrix defining the inertia tensor of the direction distribution is related to the normal form decomposition of the transfer matrix ${\bf M}_0$.

Our result on the locally averaged order parameter shows that the expression for the quadratic form of the direction inertia tensor (equation (\ref{eq: 5.1})) has the same structure as for the un-averaged case (equation (\ref{eq: 3.9})). This implies that the singularities of the averaged order parameter have the same structure as for the un-averaged case.
We also considered the Poincar\'e indices of the hyperbolic bands, where the particles become perfectly aligned. We showed that upon crossing an elliptic band with the topology of an annulus, the change of the Poincar\'e index is $\pm \frac{1}{2}$ if the trace of the transfer matrix changes sign, and $0$ if the sign of ${\rm tr}({\bf M}_0)$ is unchanged.

Finally we note that our results for recirculating flows depend upon the aspect ratio of the particles (via the parameters $\alpha_1$ and $\alpha_2=1-\alpha_1$ in (\ref{eq: 2.2})), and that in most practical applications the particles may not all have the same aspect ratio. This means that the boundaries between the elliptic and hyperbolic bands become blurred. In the ideal case, these boundaries are only marked by a discontinuity of the order parameter, so that in practical applications the boundaries may be very hard to determine. However the zeros of the order parameter are much more robust, and their normal forms have the same structure even if the rheoscopic suspension has particles which have a disperse aspect ratio.


\vspace{1.5 cm}
\noindent{\bf Acknowledgements}

VB was supported by a postgraduate fellowship from the Open University, BM is supported by the Vetenskapsr\aa{}det.

\end{document}